\documentclass[
 aip,
 jcp,
 floatfix,
 amsmath,amssymb,
 reprint
]{revtex4-1}

\usepackage{graphicx}
\usepackage{array}[=2016-10-06] % REVTeX compatibility with arXiv TeX Live 2025
\usepackage{dcolumn}
\usepackage{booktabs}
\usepackage{bm}

\usepackage[utf8]{inputenc}
\usepackage[T1]{fontenc}
\usepackage{mathptmx}
\usepackage{etoolbox}
\usepackage{algpseudocode}

\newcommand*{\VRRBL}{\mathrm{VR\mbox{-}RBL}}

\newcommand*{\dd}{\mathrm{d}}
\newcommand*{\E}{\mathbb{E}}

\newcommand*{\Cov}{\operatorname{Cov}}

\makeatletter
\def\@email#1#2{%
 \endgroup
 \patchcmd{\titleblock@produce}
  {\frontmatter@RRAPformat}
  {\frontmatter@RRAPformat{\produce@RRAP{*#1\href{mailto:#2}{#2}}}\frontmatter@RRAPformat}
  {}{}
}%
\makeatother
\begin{document}

\title[Moment-Informed Force Rescaling]{Moment-informed force rescaling for random-batch Langevin dynamics}

\author{Xingguo Wu}
\affiliation{School of Mathematical Sciences, Shanghai Jiao Tong University, Shanghai 200240, China}

\author{Yangshuai Wang}
\email{yswang@nus.edu.sg}
\affiliation{Department of Mathematics, National University of Singapore, 10 Lower Kent Ridge Road, Singapore}

\author{Zhenli Xu}
\email{xuzl@sjtu.edu.cn}
\affiliation{School of Mathematical Sciences, Shanghai Jiao Tong University, Shanghai 200240, China}
\affiliation{SOG AI-Technology Co. Ltd., Shanghai, China}
\date{16 September 2026}

\begin{abstract}
Random-batch methods accelerate molecular dynamics by replacing full interaction sums with stochastic force estimators.
In Langevin simulations, however, random-batch force errors can cause artificial heating and distort equilibrium and dynamical observables, especially for small batch sizes or under weak thermostat coupling. Moreover, the magnitude of these errors can vary across particles.
We introduce moment-informed force rescaling for random-batch list (Mi-RBL) and random-batch Ewald (Mi-RBE) methods.
A lagged average of each particle's sampled-force intensity sets an isotropic gain before the current batch is sampled, changing the random-batch force magnitude without changing its direction.
The formal finite-time analysis quantifies the bias--variance tradeoff, and an estimate of the steady-state kinetic-temperature error predicts the scaling of the kinetic error with time step and thermostat friction.
In the coexistence, binary-mixture, and electrolyte tests, Mi-RBL and Mi-RBE reduce kinetic errors and more closely reproduce the mean-square displacement, radial distribution function, and charge density profiles of the reference solutions.
Mi-RBE also reduces the growth rate of the kinetic error with the time step by more than half and retains the predicted dependence on thermostat friction.
The tested rescaling strength decreases with increasing batch size, and the particlewise update retains $O(N)$ complexity for fixed batch size.
\end{abstract}

\maketitle

\section{Introduction}

Langevin dynamics is widely used to sample canonical ensembles \cite{29acd3d494044594aea0829ef236aad6,bussi2007accurate} and to investigate the equilibrium and dynamical properties of particle systems. \cite{frenkel2001understanding,allen_computer_2017,Pastor1994} In large-scale Langevin molecular dynamics, the dominant computational cost often arises from evaluating nonbonded forces, particularly when numerous pair interactions or long-range electrostatic contributions must be included at every time step. \cite{darden_particle_1993,plimpton1995lammps,thompson2022lammps,Long-RangeelectrostaticEffects,simmonett_compression_2021}

Motivated by mini-batch and stochastic-gradient methods, the random batch method (RBM) framework reduces this cost by estimating interaction sums from small random batches. \cite{robbins1951misc-stochastic,summa_large-scale_2010,jin2020rbm,golse:hal-02405783,RBMCMB,jin_mean_2022,RB_on_guiding_problem,ye_error_2024,Cai2024ConvergenceOR} For long-range electrostatics, the random-batch Ewald (RBE) and random-batch sum-of-Gaussians methods estimate reciprocal-space contributions by random sampling. \cite{jin2021rbe,liang_superscalability_2022,liang_random_2022_npt,liang2023rbsog,quasi-2DRBE} For short-range interactions, the random-batch list (RBL) method evaluates a local core exactly and randomly samples the surrounding shell, retaining an unbiased shell-force estimator while reducing the number of pairwise evaluations. \cite{liang2021rbl,liang_improved_2022,ZHANG2026110170} Together, these approaches have enabled random-batch molecular dynamics simulations involving more than ten million particles. \cite{gao_rbmd_2025} This reduction in computational cost, however, is accompanied by stochastic force errors that enter the particle dynamics at every time step, even when the estimators are unbiased.

Unbiasedness ensures that the random-batch force error has zero conditional mean, but it does not prevent the particlewise second moments from being large and heterogeneous. In microcanonical (NVE) simulations, accumulated random-batch force errors can produce systematic energy drift, motivating energy-stable random-batch dynamics based on a weakly coupled external energy bath. \cite{liang2024ess} In canonical (NVT) Langevin simulations, the force error enters the drift term and perturbs the discrete fluctuation--dissipation balance. \cite{Kubo,xu2024vrrbl} The variance-reduced random-batch Langevin (VR-RBL) method addresses this effect by estimating the force-error covariance and modifying the Brownian noise. \cite{xu2024vrrbl} This correction acts through the thermostat rather than on the sampled force itself; the fluctuations of that force can remain strongly heterogeneous across particles and local environments. For RBL in multicomponent systems, this heterogeneity arises naturally from differences in pair parameters and local composition. \cite{liang2021rbl} Such uneven force impulses have a larger effect at small batch sizes and under weak thermostat coupling, where they can lead to artificial heating. \cite{jin_random_2020} Weak coupling is nevertheless often needed to preserve dynamical observables such as self-diffusion estimates obtained from the mean-square displacement. \cite{basconi_effects_2013}

We introduce particlewise moment-informed force rescaling to act directly on the sampled force. Treating the random-batch force as a noisy estimate of the negative potential energy gradient makes its second moment a natural quantity to monitor. Guided by second-moment normalization in stochastic optimization, we introduce a particlewise isotropic gain. \cite{leimkuhlercomputation2016,welling2011icml-bayesian,kingma2015adam,reddi2018adam} The gain is determined from a lagged exponential moving average (EMA) of the sampled-force intensity of each particle and therefore uses only information available before the current random batch is chosen. Applied to the current sampled force, the lagged gain rescales its magnitude and leaves its direction unchanged. We use the same construction for the sampled shell force in RBL and the sampled reciprocal-space force in RBE, yielding moment-informed random-batch list (Mi-RBL) and moment-informed random-batch Ewald (Mi-RBE), respectively.

We derive a formal finite-time estimate that quantifies the bias--variance tradeoff introduced by force rescaling. An approximate estimate of the steady-state kinetic-temperature error predicts the dependence of the residual kinetic error on time step and thermostat friction. In the short- and long-range tests, Mi-RBL and Mi-RBE reduce kinetic distortions and reproduce the transport, pair-structure, and screening observables of the reference more closely. Relative to RBE, Mi-RBE reduces the growth rate of the kinetic error with the time step by more than half while retaining the predicted inverse dependence on thermostat friction. Because the rescaling adds only particlewise scalar operations, Mi-RBL and Mi-RBE retain $O(N)$ complexity for fixed batch size.

\section{Method}

\subsection{Algorithmic Description}

Consider a system of $N$ particles with positions $\{\bm r_i\}_{i=1}^N$ and velocities $\{\bm v_i\}_{i=1}^N$. For a short-range pair interaction truncated at $r_s$, the RBL method partitions the neighborhood of particle $i$ into a core region $r_{ij}<r_c$ and a shell region $r_c\le r_{ij}<r_s$, where $r_c<r_s$. \cite{liang2021rbl} Let $\bm f_{ij}$ denote the force on particle $i$ exerted by particle $j$. The exact force is decomposed as the contributions from the core and shell particles:
\begin{equation}
  \bm F_i = \bm F_{i,c} + \bm F_{i,s}.
\end{equation}
RBL evaluates the core contribution $\bm F_{i,c}$ exactly and estimates the shell contribution by uniform sampling. Let $N_i^s$ be the number of shell neighbors, $P_i=\min(P,N_i^s)$ the number sampled, and $C_i$ the corresponding sampled set. For $N_i^s>0$, the shell-force estimator is
\begin{equation}
  \widetilde{\bm F}_{i,s}
  =
  \frac{N_i^s}{P_i}\sum_{j\in C_i}\bm f_{ij},
  \qquad |C_i|=P_i,
  \label{eq:rbl_shell}
\end{equation}
and $\widetilde{\bm F}_{i,s}=\bm0$ when $N_i^s=0$. Thus, all shell neighbors are used when $N_i^s\le P$. Given the current positions $\bm r$, this estimator is unbiased:
\begin{equation}
  \E\!\left[\widetilde{\bm F}_{i,s}\mid\bm r\right]
  =
  \bm F_{i,s}.
\end{equation}
The RBL force before subtracting the average force is therefore
\begin{equation}
  \widetilde{\bm F}_{i}
  =
  \bm F_{i,c}
  +
  \widetilde{\bm F}_{i,s}.
\end{equation}

To enforce zero total internal force, we subtract the average force from each particle. Equivalently, for the assembled $3N$-dimensional force vector, this operation is the projection onto the zero-total-force subspace. Define
\begin{equation}
\begin{aligned}
  \widetilde{\bm F}
  &=
  (\widetilde{\bm F}_1^T,\ldots,\widetilde{\bm F}_N^T)^T, \\
  \bm F
  &=
  (\bm F_1^T,\ldots,\bm F_N^T)^T, \\
  \mathsf P_0
  &=
  \left(\bm I_N-\frac{1}{N}\bm 1_N\bm 1_N^T\right)\otimes\bm I_3, \\
  \widetilde{\bm F}^{\,\circ}
  &=
  \mathsf P_0\widetilde{\bm F},
  \qquad
  \widetilde{\bm F}_i^{\,\circ}
  =
  \widetilde{\bm F}_i-\frac{1}{N}\sum_{j=1}^N\widetilde{\bm F}_j.
\end{aligned}
\label{eq:zero_force_projection}
\end{equation}
Here $\bm I_N$ and $\bm I_3$ are identity matrices, $\bm 1_N$ is the $N$-component vector of ones, and $\otimes$ denotes the Kronecker product. Because the exact internal force satisfies $\sum_i\bm F_i=\bm0$, we have $\mathsf P_0\bm F=\bm F$. Define the raw sampling error $\bm\zeta=\widetilde{\bm F}-\bm F$ and the projected force error $\bm\eta=\widetilde{\bm F}^{\,\circ}-\bm F$. The conditional unbiasedness of the shell-force estimator gives
\begin{equation}
\begin{aligned}
  \bm\eta
  &=
  \mathsf P_0\bm\zeta, \\
  \E[\bm\eta\mid\bm r]
  &=
  \bm0,
  \qquad
  \sum_{i=1}^N\bm\eta_i=\bm0.
\end{aligned}
\label{eq:rbl_projected_error}
\end{equation}
The RBL Langevin dynamics can therefore be written as
\begin{equation}
\left\{
\begin{aligned}
\dd \bm r_i
&= \bm v_i\,\dd t, \\[3pt]
m_i\,\dd \bm v_i
&=
\left[
\bm F_i(\bm r)
+\bm \eta_i
-\gamma m_i\bm v_i
\right]\dd t
+\sqrt{2\gamma k_B Tm_i}\,\dd\bm W_i ,
\end{aligned}
\right.
\label{RBLLangevin}
\end{equation}
Here $m_i$ is the particle mass, $\gamma$ is the thermostat friction coefficient, $k_B$ is the Boltzmann constant, $T$ is the target temperature, and $\{\bm W_i\}_{i=1}^{N}$ are independent three-dimensional Wiener processes. 

Because the core force is evaluated exactly, the raw error $\bm\zeta_i$ defined above arises entirely from shell sampling. For a fixed configuration, write $\widetilde{\bm F}_{i,s}=\bm F_{i,s}+\bm\zeta_i$ with $\E[\bm\zeta_i\mid\bm r]=\bm0$. Then
\begin{equation}
  \E\!\left[\|\widetilde{\bm F}_{i,s}\|^2\mid\bm r\right]
  =
  \|\bm F_{i,s}\|^2
  +
  \operatorname{Tr}\Cov(\bm\zeta_i\mid\bm r).
  \label{eq:sampled_force_second_moment}
\end{equation}
Thus, although conditional unbiasedness fixes the mean, the covariance term increases the sampled-force second moment and can vary strongly across particles. In multicomponent or spatially heterogeneous systems, differences in pair parameters, local composition, and shell-neighbor configurations make this noise intensity nonuniform. Mi-RBL stores a lagged exponential moving average of $\|\widetilde{\bm F}_{i,s}\|^2/3$ and uses it to assign smaller gains to particles with larger lagged force intensities.

Mi-RBL applies the rescaling to the sampled shell-force component. We initialize the particlewise moment state with $s_{i,0}=0$. At the beginning of step $n$, $s_{i,n}$ contains the sampled-force information accumulated through step $n-1$. Let $\mathcal F_n$ denote the information available immediately before the shell batch at step $n$ is sampled. For $n\ge1$, the lagged state is bias corrected as
\begin{equation}
  \widehat s_{i,n}
  =
  \frac{s_{i,n}}{1-\beta^n},
  \label{eq:mirbl_state_update2}
\end{equation}
and used to form the gain
\begin{equation}
  g_{i,n}(\alpha)
  =
  \left(
    \frac{\widehat s_{i,n}}{S_0}+\varepsilon_\ast
  \right)^{-\alpha/2},
  \label{eq:mirbl_gain}
\end{equation}
where $0<\beta<1$ is the EMA parameter, $S_0$ is a fixed force-squared reference scale, $\varepsilon_\ast>0$ is a dimensionless stabilizer, and $\alpha\ge0$ controls the rescaling strength. Because no lagged state is available initially, we set $g_{i,0}=1$. All calculations use $\varepsilon_\ast=10^{-8}$.

The gains are therefore fixed before the current shell batch is sampled. After sampling $\widetilde{\bm F}_{i,s}^{\,n}$, Mi-RBL constructs
\begin{equation}
  \widehat{\bm F}_{i,s}^{\,n}
  =
  g_{i,n}(\alpha)\widetilde{\bm F}_{i,s}^{\,n}.
  \label{eq:mirbl_transform}
\end{equation}
Since $g_{i,n}$ is a positive scalar, this rescaling preserves the sampled-force direction and is covariant under rotations. Define the gain matrix
\begin{equation}
  \bm G_n
  =
  \operatorname{diag}
  \left(
    g_{1,n}\bm I_3,\ldots,g_{N,n}\bm I_3
  \right),
  \label{eq:mirbl_gain_matrix}
\end{equation}
Since the matrix $\bm G_n$ is determined before the current batch is sampled, it is $\mathcal F_n$-measurable. Using $\E[\bm\zeta_n\mid\mathcal F_n]=\bm0$ for the current raw sampling error, we have
\begin{equation}
  \E\left[
    \mathsf P_0\bm G_n\bm\zeta_n
    \mid\mathcal F_n
  \right]
  =\mathsf P_0\bm G_n\E\left[
    \bm\zeta_n
    \mid\mathcal F_n
  \right]
  =
  \bm0.
  \label{eq:mirbl_centered_noise}
\end{equation}
This is the reason for using the lagged gain: $\bm G_n$ contains only historical information and no information from the current random batch, so the first equality above is valid, thereby keeping the transformed random-batch fluctuation conditionally centered. The final zero-total-force Mi-RBL force is
\begin{equation}
  \widehat{\bm F}^{\,n}
  =
  \mathsf P_0
  \left(
    \bm F_c^{\,n}
    +
    \bm G_n\widetilde{\bm F}_s^{\,n}
  \right), 
  \label{eq:mirbl_combinedforce}
\end{equation}
where $\bm F_c^{\,n}$ and $\widetilde{\bm F}_s^{\,n}$ are the stacked core and sampled shell forces. When $\alpha=0$, $\bm G_n=\bm I$, and the construction reduces to RBL.

After the sampled shell force $\widetilde{\bm F}_{i,s}^{\,n}$ has been used in Eq.~\eqref{eq:mirbl_combinedforce}, it is used to update the moment state:
\begin{equation}
  s_{i,n+1}
  =
  \beta s_{i,n}
  +
  (1-\beta)\frac{\|\widetilde{\bm F}_{i,s}^{\,n}\|^2}{3}.
  \label{eq:mirbl_state_update1}
\end{equation}
Here $\beta$ sets the memory of the EMA. At the beginning of step $n+1$, $s_{i,n+1}$ is divided by $1-\beta^{n+1}$, as in Eq.~\eqref{eq:mirbl_state_update2}. Since $g_{i,n}$ is computed from $s_{i,n}$ before the step-$n$ shell batch is sampled, the sampled force at step $n$ can affect $g_{i,n+1}$ through $s_{i,n+1}$, but it cannot affect $g_{i,n}$. Writing $\widetilde{\bm F}_{i,s}^{\,n}=\bm F_{i,s}^{\,n}+\bm\zeta_{i,n}$, the rescaled shell-force second moment satisfies
\begin{equation}
\begin{aligned}
  &\E\!\left[
    \|g_{i,n}\widetilde{\bm F}_{i,s}^{\,n}\|^2
    \mid\mathcal F_n
  \right] \\
  &\quad =
  g_{i,n}^2
  \left(
    \|\bm F_{i,s}^{\,n}\|^2
    +
    2\bm F_{i,s}^{\,n}\cdot
    \E[\bm\zeta_{i,n}\mid\mathcal F_n]
    +
    \E[\|\bm\zeta_{i,n}\|^2\mid\mathcal F_n]
  \right) \\
  &\quad =
  g_{i,n}^2
  \left(
    \|\bm F_{i,s}^{\,n}\|^2
    +
    \operatorname{Tr}\bm\Sigma_{ii,n}
  \right),
\end{aligned}
  \label{eq:mirbl_rescaled_second_moment}
\end{equation}
where $\bm\Sigma_{ii,n}:=\Cov(\bm\zeta_{i,n}\mid\mathcal F_n)$. The first equality uses that $g_{i,n}$ and $\bm F_{i,s}^{\,n}$ are $\mathcal F_n$-measurable. The second uses $\E[\bm\zeta_{i,n}\mid\mathcal F_n]=\bm0$. A large $s_{i,n+1}$ gives a smaller gain at the next step. When $g_{i,n+1}<1$, the corresponding variance term is reduced by the factor $g_{i,n+1}^2$. Since the state is updated separately for each particle, the strength of suppression can adjust during the simulation for particles of different species and local environments. A small $s_{i,n+1}$ may instead give $g_{i,n+1}>1$ and allow low-intensity sampled shell forces to recover. However, the same scaling also biases the conditional mean of the sampled shell force, changing it from $\bm F_{i,s}^{\,n}$ to $g_{i,n}\bm F_{i,s}^{\,n}$. This is the bias--variance tradeoff introduced in the following error analysis.
\paragraph{Extension to random-batch Ewald.}
For systems with long-range Coulomb interactions, we write the force as
\begin{equation}
  \bm F_i
  =
  \bm F_{i,\mathrm{real}}
  +
  \bm F_{i,\mathrm{rec}},
\end{equation}
where the real-space contribution is evaluated with a short-range cutoff and the reciprocal-space contribution is given by the Ewald Fourier sum. \cite{ewald1921berechnung,hu2014infinite}
Let $\alpha_E$ denote the Ewald splitting parameter. The reciprocal-space force is
\begin{equation}
\begin{aligned}
\bm F_{i,\mathrm{rec}}
&=
-\sum_{\bm k\ne\bm0}
\frac{4\pi q_i\bm k}{V|\bm k|^2}
\exp\left(-\frac{|\bm k|^2}{4\alpha_E}\right)
\operatorname{Im}\left[
e^{-i\bm k\cdot\bm r_i}\rho(\bm k)
\right], \\[4pt]
\rho(\bm k)
&=
\sum_{j=1}^{N}q_j e^{i\bm k\cdot\bm r_j}.
\end{aligned}
\label{eq:ewald_rec_force}
\end{equation}
The particle--particle particle--mesh (PPPM) method evaluates this reciprocal-space contribution through mesh interpolation and fast Fourier transforms (FFTs) with $O(N\log N)$ complexity. \cite{hockney1988computer,darden_particle_1993,essmann_smooth_1995}

RBE replaces the full reciprocal sum by importance sampling. \cite{jin2021rbe}
Define
\begin{equation}
  S
  =
  \sum_{\bm k\ne\bm0}
  \exp\left(-\frac{|\bm k|^2}{4\alpha_E}\right),
  \qquad
  p(\bm k)
  =
  S^{-1}
  \exp\left(-\frac{|\bm k|^2}{4\alpha_E}\right).
\end{equation}
At each time step, $P$ reciprocal modes $\{\bm k_l\}_{l=1}^{P}$ are sampled from $p(\bm k)$, giving
\begin{equation}
  \widetilde{\bm F}_{i,\mathrm{rec}}
  =
  -\frac{S}{P}
  \sum_{l=1}^P
  \frac{4\pi q_i\bm k_l}{V|\bm k_l|^2}
  \operatorname{Im}\left[
    e^{-i\bm k_l\cdot\bm r_i}\rho(\bm k_l)
  \right].
  \label{eq:rbe_force}
\end{equation}
The same modes are used for all particles, allowing the corresponding structure factors to be shared. For fixed $P$, evaluating the sampled reciprocal-space contribution requires $O(N)$ operations. \cite{jin2021rbe,liang_superscalability_2022}
For the corresponding $3N$-dimensional force vectors, define the raw RBE error at step $n$ by
\begin{equation}
  \bm\zeta_{\mathrm{rec},n}
  =
  \widetilde{\bm F}_{\mathrm{rec}}^{\,n}
  -
  \bm F_{\mathrm{rec}}^{\,n}.
\end{equation}
Conditional on $\mathcal F_n$, the estimator is unbiased:
\begin{equation}
  \E\left[
    \bm\zeta_{\mathrm{rec},n}
    \mid \mathcal F_n
  \right]
  =
  \bm0.
  \label{eq:rbe_conditional_mean}
\end{equation}
Because the same reciprocal modes are used for all particles, the force errors of different particles are generally correlated. The analysis below therefore keeps the full conditional covariance matrix.

Mi-RBE reuses the lagged Mi-RBL construction, with the sampled shell force replaced by the RBE reciprocal-force estimator and the exact core force replaced by the real-space force. At the beginning of step $n$, the lagged state $s_{i,n}$ is bias corrected using Eq.~\eqref{eq:mirbl_state_update2}, and the gain $g_{i,n}$ is formed using Eq.~\eqref{eq:mirbl_gain}; the same initialization $g_{i,0}=1$ applies. The gain is therefore fixed before the current reciprocal modes are sampled. Once $\widetilde{\bm F}_{i,\mathrm{rec}}^{\,n}$ has been evaluated, Mi-RBE constructs
\begin{equation}
  \widehat{\bm F}_{i,\mathrm{rec}}^{\,n}
  =
  g_{i,n}(\alpha)
  \widetilde{\bm F}_{i,\mathrm{rec}}^{\,n}.
  \label{eq:mirbe_transform}
\end{equation}
Here $\bm G_n$ is the gain matrix in Eq.~\eqref{eq:mirbl_gain_matrix}. Since $\bm G_n$ is fixed before the reciprocal modes are sampled and $\E[\bm\zeta_{\mathrm{rec},n}\mid\mathcal F_n]=\bm0$, the transformed reciprocal-space random-batch fluctuation has zero conditional mean:
\begin{equation}
  \E\left[
    \mathsf P_0
    \bm G_n
    \bm\zeta_{\mathrm{rec},n}
    \mid \mathcal F_n
  \right]
  =
  \bm0.
  \label{eq:mirbe_centered_noise}
\end{equation}
The final Mi-RBE force is projected as
\begin{equation}
  \widehat{\bm F}^{\,n}
  =
  \mathsf P_0
  \left(
    \bm F_{\mathrm{real}}^{\,n}
    +
    \bm G_n
    \widetilde{\bm F}_{\mathrm{rec}}^{\,n}
  \right).
  \label{eq:mirbe_combinedforce}
\end{equation}
This projection enforces zero total internal force after the particlewise rescaling. When $\alpha=0$, $\bm G_n=\bm I$, and the construction reduces to RBE.

Only after the step-$n$ force has been constructed is the reciprocal-force moment state updated:
\begin{equation}
  s_{i,n+1}
  =
  \beta s_{i,n}
  +
  (1-\beta)
  \frac{
    \|\widetilde{\bm F}_{i,\mathrm{rec}}^{\,n}\|^2
  }{3}.
  \label{eq:mirbe_state_update}
\end{equation}
Thus, the reciprocal batch sampled at step $n$ can affect $g_{i,n+1}$ but not $g_{i,n}$. At fixed $P$, Mi-RBE retains the $O(N)$ reciprocal-space complexity of RBE and adds only one scalar state update and one scalar multiplication per particle. The lagged gain construction is summarized in Algorithm~1. In all numerical comparisons, it is embedded in the same velocity-Verlet Langevin integrator as the corresponding baseline method.

\par\medskip
\noindent\rule{\linewidth}{0.7pt}\\[-0.8ex]
\noindent\rule{\linewidth}{0.3pt}
\vspace{0.1em}
\noindent\textbf{Algorithm 1. Lagged moment-informed random-batch force construction at step $n$.}
\begin{algorithmic}[1]
\Require Positions $\bm r^n$, lagged states $\{s_{i,n}\}_{i=1}^N$, batch size $P$, and parameters $\alpha$, $\beta$, $S_0$, and $\varepsilon_\ast$
\State Compute $\bm F_c^{\,n}$ for Mi-RBL, or $\bm F_{\mathrm{real}}^{\,n}$ for Mi-RBE.
\State For $n=0$, set $g_{i,0}=1$ for all particles; for $n\ge1$, compute the lagged gains $\{g_{i,n}\}_{i=1}^N$ using Eqs.~\eqref{eq:mirbl_state_update2} and \eqref{eq:mirbl_gain}.
\State Sample $\widetilde{\bm F}_s^{\,n}$ using Eq.~\eqref{eq:rbl_shell} for Mi-RBL, or $\widetilde{\bm F}_{\mathrm{rec}}^{\,n}$ using Eq.~\eqref{eq:rbe_force} for Mi-RBE.
\State Rescale the random-batch force estimate using Eq.~\eqref{eq:mirbl_transform} for Mi-RBL, or Eq.~\eqref{eq:mirbe_transform} for Mi-RBE.
\State Subtract the average force, equivalently applying $\mathsf P_0$, to obtain $\widehat{\bm F}^{\,n}$ from Eq.~\eqref{eq:mirbl_combinedforce} for Mi-RBL or Eq.~\eqref{eq:mirbe_combinedforce} for Mi-RBE.
\State Update $\{s_{i,n+1}\}_{i=1}^N$ using Eq.~\eqref{eq:mirbl_state_update1} for Mi-RBL, or Eq.~\eqref{eq:mirbe_state_update} for Mi-RBE.
\State \Return $\widehat{\bm F}^{\,n}$ and $\{s_{i,n+1}\}_{i=1}^N$.
\end{algorithmic}
\vspace{0.1em}
\noindent\rule{\linewidth}{0.3pt}\\[-0.8ex]
\noindent\rule{\linewidth}{0.7pt}
\par\medskip

\subsection{Error Analysis}

We use the same force-error decomposition for Mi-RBL and Mi-RBE. It yields two results: a formal finite-time bound on the distributional deviation and an approximate estimate of the steady-state kinetic-temperature error.

For the system of $N$ particles, let
\begin{equation}
  \bm r
  =
  (\bm r_1^T,\ldots,\bm r_N^T)^T\in\mathbb R^{3N},
  \qquad
  \bm v
  =
  (\bm v_1^T,\ldots,\bm v_N^T)^T\in\mathbb R^{3N}.
\end{equation}
Write the exact internal force as
\begin{equation}
  \bm F
  =
  \bm F_a+\bm F_b,
\end{equation}
where $\bm F_b$ is the component approximated by random-batch sampling and $\bm F_a$ is evaluated directly. Thus,
\begin{equation}
\begin{aligned}
  (\bm F_a,\bm F_b)
  &=
  (\bm F_c,\bm F_s)
  &&\text{for Mi-RBL}, \\
  (\bm F_a,\bm F_b)
  &=
  (\bm F_{\mathrm{real}},\bm F_{\mathrm{rec}})
  &&\text{for Mi-RBE}.
\end{aligned}
\end{equation}

At step $n$, write the random-batch estimator as
\begin{equation}
\begin{aligned}
  \widetilde{\bm F}_{b,n}
  &=
  \bm F_b(\bm r_n)+\bm\zeta_n, \\
  \E[\bm\zeta_n\mid\mathcal F_n]
  &=
  \bm0, \\
  \Cov(\bm\zeta_n\mid\mathcal F_n)
  &=
  \bm\Sigma_n.
\end{aligned}
  \label{eq:generic_batch_error}
\end{equation}
Here $\mathcal F_n$ denotes the information available at the start of step $n$, including the current phase-space state and the EMA states $\{s_{i,n}\}_{i=1}^N$ accumulated from earlier batches. Hence $\bm G_n$ is $\mathcal F_n$-measurable.

At step $n$, the rescaled scheme uses
\begin{equation}
  \widehat{\bm F}_n
  =
  \mathsf P_0
  \left[
    \bm F_a(\bm r_n)
    +
    \bm G_n\widetilde{\bm F}_{b,n}
  \right].
\end{equation}
Since the exact internal force satisfies $\mathsf P_0\bm F=\bm F$, the force error is
\begin{equation}
\begin{aligned}
  \bm e_n
  &:={}
  \widehat{\bm F}_n-\bm F(\bm r_n) \\
  &=
  \underbrace{
    \mathsf P_0
    (\bm G_n-\bm I)
    \bm F_b(\bm r_n)
  }_{\bm b_n}
  +
  \underbrace{
    \mathsf P_0
    \bm G_n\bm\zeta_n
  }_{\bm\xi_n}.
\end{aligned}
\label{eq:error_decomp}
\end{equation}
The first term is the force-scaling bias, and the second is the transformed random-batch error. The lagged construction gives
\begin{equation}
  \E[\bm\xi_n\mid\mathcal F_n]
  =
  \bm0,
  \qquad
  \Cov(\bm\xi_n\mid\mathcal F_n)
  =
  \bm Q_n(\alpha),
\end{equation}
where
\begin{equation}
  \bm Q_n(\alpha)
  =
  \mathsf P_0
  \bm G_n
  \bm\Sigma_n
  \bm G_n^T
  \mathsf P_0^T.
  \label{eq:transformed_covariance}
\end{equation}
If the zero-total-force projection is omitted, the covariance part has the particlewise scaling shown in Eq.~\eqref{eq:mirbl_rescaled_second_moment}. In particular, the particle-$i$ covariance block scales as
\begin{equation}
  \bm\Sigma_{ii,n}\mapsto g_{i,n}^2\bm\Sigma_{ii,n},
  \qquad
  \operatorname{Tr}\bm\Sigma_{ii,n}\mapsto
  g_{i,n}^2\operatorname{Tr}\bm\Sigma_{ii,n}.
\end{equation}
When $g_{i,n}<1$, this scaling reduces $\operatorname{Tr}\bm\Sigma_{ii,n}$.

We next expand the bias term in Eq.~\eqref{eq:error_decomp}. Set $\bm L_0=\bm0$ and, for $n\ge1$, define
\begin{equation}
\begin{aligned}
  \ell_{i,n}
  &=
  \log\left(
    \frac{\widehat s_{i,n}}{S_0}
    +
    \varepsilon_\ast
  \right), \\
  \bm L_n
  &=
  \operatorname{diag}
  (\ell_{1,n}\bm I_3,\ldots,\ell_{N,n}\bm I_3).
\end{aligned}
\end{equation}
For small $\alpha$,
\begin{equation}
  \bm G_n
  =
  \bm I
  -
  \frac{\alpha}{2}\bm L_n
  +
  O(\alpha^2),
\end{equation}
and therefore
\begin{equation}
  \bm b_n
  =
  -\frac{\alpha}{2}
  \mathsf P_0
  \bm L_n
  \bm F_b(\bm r_n)
  +
  O(\alpha^2).
  \label{eq:bias_expansion}
\end{equation}
If $N^{-1}\|\bm F_b(\bm r_n)\|^2$ and $\|\bm L_n\|$ are uniformly bounded, this expansion gives
\begin{equation}
  \frac{1}{N}
  \|\bm b_n\|^2
  =
  O(\alpha^2).
\end{equation}

For the accumulated contributions of the bias $\bm b_n$ and covariance $\bm Q_n(\alpha)$, let $t_n=n\Delta t$, $t^*=N_{t^*}\Delta t$, and define
\begin{equation}
\begin{aligned}
  \mathcal B_{t^*,N}(\alpha)
  :=
  &\sup_R
  \frac{1}{Nt^*}
  \sum_{n=0}^{N_{t^*}-1}
  \Delta t\,
  \E_R\|\bm b_n\|^2, \\
  \mathcal Q_{t^*,N}(\alpha)
  :=
  &\sup_R
  \frac{1}{Nt^*}
  \sum_{n=0}^{N_{t^*}-1}
  \Delta t\,
  \E_R\operatorname{Tr}\bm Q_n(\alpha).
\end{aligned}
\label{eq:accumulated_error_terms}
\end{equation}
Here, $\E_R$ denotes the expectation conditional on the initial phase-space state $R$ and averages over both thermostat and random-batch realizations. 

Starting from the same initial state $R$, let $\mu_{t^*}^R$ and $\widehat\mu_{t^*}^{\alpha,R}$ denote the phase-space distributions at time $t^*$ of the exact Langevin dynamics and an auxiliary dynamics with rescaled random-batch forces, respectively. We measure their finite-time difference by the normalized 2-Wasserstein distance \cite{villani2009optimal}
\begin{equation}
  W_{2,N}(\mu,\nu)
  =
  \left[
    \inf_{\lambda\in\Pi(\mu,\nu)}
    \int_{\mathbb R^{6N}\times\mathbb R^{6N}}
    \frac{1}{N}
    \|\mathbf x-\mathbf y\|_2^2
    \,\dd\lambda(\mathbf x,\mathbf y)
  \right]^{1/2},
  \label{eq:wasserstein_distance}
\end{equation}
where $\Pi(\mu,\nu)$ is the set of couplings with marginals $\mu$ and $\nu$.

\noindent\textbf{Formal finite-time estimate.}
Assume that $\bm F$ is uniformly Lipschitz, $\bm F_b$ has a bounded per-particle second moment, and the particle masses and logarithmic factors in $\bm L_n$ are uniformly bounded. Then, for any finite $t^*>0$, there exist $\alpha_0,\Delta t_0>0$ and a constant $C(t^*)$, independent of $N$, $\alpha$, and $\Delta t$, such that, for $0\le\alpha\le\alpha_0$ and $0<\Delta t\le\Delta t_0$,
\begin{equation}
\begin{aligned}
  &\sup_R W_{2,N}\left(
    \mu_{t^*}^R,
    \widehat\mu_{t^*}^{\alpha,R}
  \right) \\
  &\qquad\le
  C(t^*)
  \sqrt{
    \mathcal B_{t^*,N}(\alpha)
    +
    \Delta t\,\mathcal Q_{t^*,N}(\alpha)
  }.
\end{aligned}
  \label{eq:tradeoff}
\end{equation}
The expansion in Eq.~\eqref{eq:bias_expansion} gives $\mathcal B_{t^*,N}(\alpha)=O(\alpha^2)$, and Eq.~\eqref{eq:tradeoff} bounds the finite-time distributional error by the accumulated force-scaling bias and the transformed random-batch covariance. The proof is given in Appendix~A. This estimate is obtained under the smoothness and boundedness assumptions introduced above. The unregularized Lennard--Jones and Coulomb forces used in the numerical tests are singular, and therefore fall outside the direct scope of the estimate. Nevertheless, the estimate indicates how force rescaling changes the balance between the bias and covariance contributions in random-batch dynamics.

When the effective dynamics has reached a steady state, we connect the force error to the kinetic statistics through a continuous-time approximation. The random-batch force error over one step is replaced by a Brownian term with the same covariance. For simplicity, all particle masses are taken to be $m$, and the lagged gain is treated as fixed during one time step. Thus,
\begin{equation}
  2\bm D_\alpha\Delta t
  =
  \Delta t^2\bm Q_\alpha.
\end{equation}
Here, $\Delta t^2\bm Q_\alpha$ is the covariance accumulated by the random-batch force error over one step, and $2\bm D_\alpha\Delta t$ is the covariance of the matched Brownian increment. In the following, $\bm b_\alpha$ and $\bm Q_\alpha$ denote the steady-state bias and transformed covariance terms defined in Eqs.~\eqref{eq:error_decomp} and \eqref{eq:transformed_covariance}. $\E_{\mathrm{ss}}$ denotes expectation in the stationary state of the effective dynamics. We write the per-particle steady-state averages as
\begin{equation}
  \langle\bm v\cdot\bm b_\alpha\rangle
  :=
  \frac{1}{N}\E_{\mathrm{ss}}[\bm v^T\bm b_\alpha],
  \qquad
  \langle\operatorname{Tr}\bm Q_\alpha\rangle
  :=
  \frac{1}{N}\E_{\mathrm{ss}}[\operatorname{Tr}\bm Q_\alpha].
\end{equation}
Using the covariance matching above, we obtain
\begin{equation}
  T_\alpha-T
  \approx
  \frac{1}{3\gamma k_B}
  \langle\bm v\cdot\bm b_\alpha\rangle
  +
  \frac{\Delta t}{6\gamma m k_B}
  \langle\operatorname{Tr}\bm Q_\alpha\rangle.
  \label{eq:temp_balance_general}
\end{equation}
The derivation is given in Appendix~B. Here $T$ is the prescribed thermostat temperature, and $T_\alpha$ denotes the stationary kinetic temperature of the effective dynamics with force rescaling strength $\alpha$. The term $\langle\bm v\cdot\bm b_\alpha\rangle$ is the velocity-bias correlation and may have either sign. The second term is nonnegative and is determined by the transformed random-batch covariance. The formula shows that force rescaling affects the kinetic shift through two quantities: the covariance term $\langle\operatorname{Tr}\bm Q_\alpha\rangle$ and the bias term $\langle\bm v\cdot\bm b_\alpha\rangle$.

For $\alpha=0$, the bias vanishes and $\bm Q_0=\mathsf P_0\bm\Sigma\mathsf P_0^T$, so Eq.~\eqref{eq:temp_balance_general} reduces to
\begin{equation}
  T_0-T
  \approx
  \frac{\Delta t}{6\gamma m k_B}
  \langle\operatorname{Tr}\bm Q_0\rangle.
  \label{eq:rbl_heating}
\end{equation}
Thus, for RBL and RBE, the estimate predicts a positive temperature shift caused by the random-batch covariance.

If the bias and covariance averages depend weakly on $\gamma$, the kinetic error scales approximately as $\gamma^{-1}$. At fixed $\gamma$, if $\bm Q_\alpha$ varies little with $\Delta t$, the covariance contribution is $O(\Delta t)$. Hence, if
\begin{equation}
  \langle\operatorname{Tr}\bm Q_\alpha\rangle
  <
  \langle\operatorname{Tr}\bm Q_0\rangle,
\end{equation}
then the rescaled dynamics can reduce the kinetic error for suitable $\alpha$, as long as the bias term does not dominate.

The VR-RBL correction used in the RBL tests follows Ref.~\citenum{xu2024vrrbl} and is summarized in Appendix~B.

\section{Results}

For the short-range benchmarks, we compare classical Langevin dynamics (CL), RBL, VR-RBL, and Mi-RBL. CL computes all pair forces within the prescribed cutoff and serves as the reference calculation. For the long-range benchmark, PPPM provides the reference for RBE and Mi-RBE. Within each system, all methods use the same target temperature, time step, isotropic Langevin friction $\bm\Gamma=\gamma\bm I$, and velocity-Verlet integration settings.

VR-RBL is included in the short-range comparisons. Following Ref.~\citenum{xu2024vrrbl}, the shell force is first sampled as in Eq.~\eqref{eq:rbl_shell}. The covariance of the discrete Brownian increment is then modified by subtracting an estimate of the random-batch force covariance. For particle $i$, this gives
\begin{equation}
  2\gamma m_i k_B T\Delta t\,\bm I_3
  -
  \Delta t^2\widehat{\bm\Sigma}_i,
  \label{eq:vrrbl_brownian_covariance}
\end{equation}
where $\widehat{\bm\Sigma}_i$ estimates the covariance of the raw shell-sampling error $\bm\zeta_i$. Because shell neighbors are sampled without replacement in the present work, $\widehat{\bm\Sigma}_i$ is computed with the finite-population correction given in Appendix~B, rather than with the estimator of Ref.~\citenum{xu2024vrrbl}. If the matrix in Eq.~\eqref{eq:vrrbl_brownian_covariance} has a negative eigenvalue, the Brownian increment is omitted for that particle and time step. This is the fallback used in the present calculations. 

The reported observables characterize different effects of random-batch fluctuations. The kinetic mean indicates systematic heating or cooling, and the kinetic standard deviation $\sigma_k$ measures temporal kinetic fluctuations. The mean-square displacement (MSD) probes transport, the radial distribution functions (RDFs) characterize pair structure, and the cation-centered charge density measures ionic screening. For the per-atom kinetic energy $k(t)=K_{\mathrm{kin}}(t)/N$, where $K_{\mathrm{kin}}(t)$ is the total kinetic energy, we define
\begin{equation}
  \bar{k}=\langle k(t)\rangle_t,
  \qquad
  \sigma_k=
  \left\langle
    [k(t)-\bar{k}]^2
  \right\rangle_t^{1/2}.
\end{equation}
The MSD is evaluated as
\begin{equation}
  \operatorname{MSD}(\ell)
  =
  \left\langle
    \left|
      \bm r_i(t+\ell)-\bm r_i(t)
    \right|^2
  \right\rangle_{i,t},
\end{equation}
where the average is taken over particles and multiple time origins after removing the drift of the mass center. Reported statistics are averaged over eight independent trajectories. Production runs were performed on a Linux Slurm cluster equipped with Intel Xeon Platinum 8358 processors at $2.60$ GHz.

\subsection{Gas--Liquid Coexistence Lennard-Jones System}

The gas--liquid coexistence test examines whether Mi-RBL preserves long-time transport under weak thermostat coupling. We use a Lennard-Jones fluid with $N=7000$, number density $\rho=0.2$, and temperature $T=0.9$. \cite{verlet1967computer,watanabe2012phase} Lennard-Jones reduced units are defined by $\sigma_{\mathrm{LJ}}=\epsilon_{\mathrm{LJ}}=m=k_B=1$ and $\tau_{\mathrm{LJ}}=\sigma_{\mathrm{LJ}}\sqrt{m/\epsilon_{\mathrm{LJ}}}$. The core and shell cutoffs are $r_c=2.0\sigma_{\mathrm{LJ}}$ and $r_s=6.0\sigma_{\mathrm{LJ}}$, respectively. All simulations use $\Delta t=0.01\tau_{\mathrm{LJ}}$ and $\bm\Gamma=0.25\bm I$. For Mi-RBL, $S_0=(\epsilon_{\mathrm{LJ}}/\sigma_{\mathrm{LJ}})^2=1$. Each trajectory is equilibrated for $2\times10^5$ steps, followed by $3\times10^5$ production steps.

RBL, VR-RBL, and Mi-RBL use shell batch sizes $P=5$, $10$, and $15$, while Mi-RBL fixes $\beta=0.95$. To keep parameter selection separate from the production runs, we used short pilot runs with four random seeds to identify the tested $\alpha$ intervals from equilibrated kinetic statistics and short-trajectory MSDs. These intervals were then fixed before the final production runs. The selected ranges are $0.58\le\alpha\le0.60$ for $P=5$, $0.36\le\alpha\le0.38$ for $P=10$, and $0.26\le\alpha\le0.28$ for $P=15$. Their downward shift with increasing $P$ agrees with the weaker random-batch fluctuations at larger batch sizes, which require less rescaling.

\begin{figure*}[htbp]
  \centering
  \includegraphics[width=0.97\textwidth]{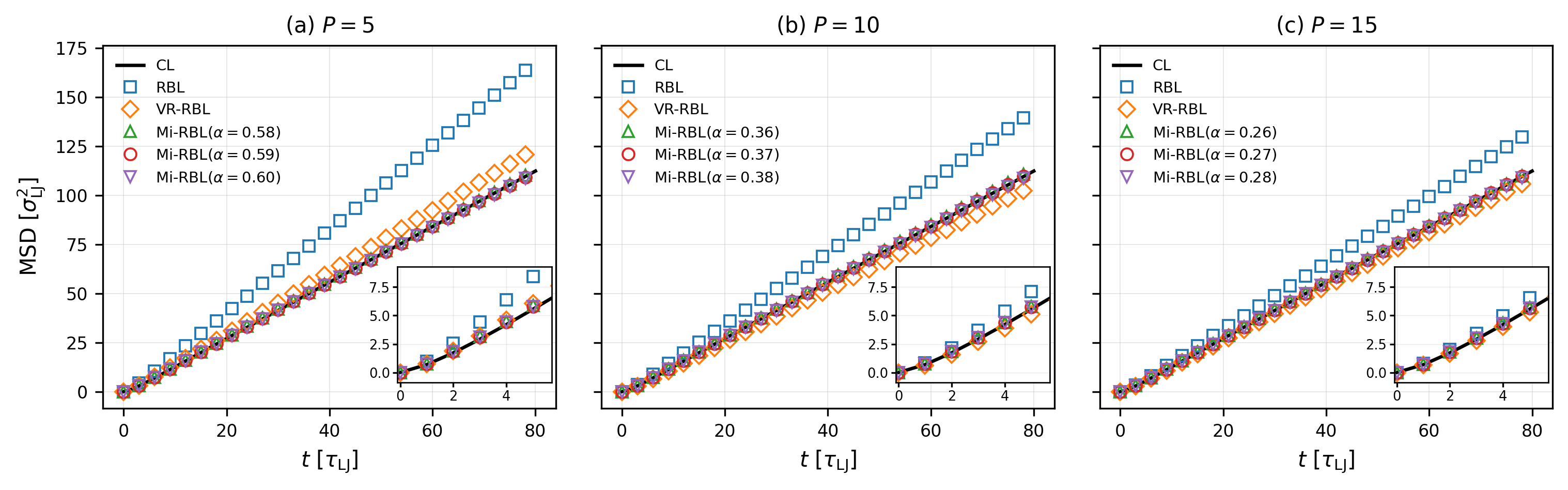}
  \caption[Gas-liquid coexistence LJ MSD comparisons]{Mean-square displacement of the gas--liquid coexistence Lennard-Jones system as a function of time for (a) $P=5$, (b) $P=10$, and (c) $P=15$. The Mi-RBL values of $\alpha$ are $0.58$, $0.59$, and $0.60$ in (a); $0.36$, $0.37$, and $0.38$ in (b); and $0.26$, $0.27$, and $0.28$ in (c), with $\beta=0.95$. The insets enlarge the short-time region.}
  \label{fig:coex_msd}
\end{figure*}

Figure~\ref{fig:coex_msd} shows the usual ballistic-to-diffusive crossover: the MSD is approximately quadratic at short times and becomes linear at long times. RBL overestimates the long-time displacement, with the largest deviation at $P=5$. VR-RBL reduces this deviation but remains visibly different from CL, particularly at $P=5$. At all three batch sizes and for every tested $\alpha$, Mi-RBL is closer to the CL reference.

\subsection{Kob--Andersen Binary Lennard-Jones Mixture}

A heterogeneous Kob--Andersen $80:20$ binary Lennard-Jones mixture is used to test whether Mi-RBL controls kinetic and pair-structural distortions in a multicomponent system. \cite{kob1995testing,malins_lifetimes_2013} Because the pair parameters and local composition vary with species, the sampled shell-force intensity is nonuniform among particles. This makes the system a useful test for particlewise rescaling. The system contains $N=8788$ particles at number density $\rho=1.20$ and target temperature $T=0.60$. Both species have mass $m=1$, and we set $k_B=1$. The pair parameters and interaction cutoffs are listed in Table~\ref{tbl:binary_lj_params}.
\begin{table}[htbp]
  \caption{Kob--Andersen binary Lennard-Jones pair parameters and pair-specific interaction cutoffs. Energies are normalized by $\epsilon_{11}$, whereas lengths and cutoffs are normalized by $\sigma_{11}$.}
  \label{tbl:binary_lj_params}
  \centering
  \begin{ruledtabular}
  \begin{tabular}{cccc}
    
    Pair & $\epsilon_{ab}/\epsilon_{11}$ & $\sigma_{ab}/\sigma_{11}$ & $r_{\mathrm{cut},ab}/\sigma_{11}$ \\
    \hline
    $1$--$1$ & $1.0$ & $1.00$ & $2.50$ \\
    $1$--$2$ & $1.5$ & $0.80$ & $2.00$ \\
    $2$--$2$ & $0.5$ & $0.88$ & $2.20$ \\
    
  \end{tabular}
  \end{ruledtabular}
\end{table}
RBL uses a common shell cutoff $r_s=2.50\sigma_{11}$, which is the largest pair-specific cutoff in Table~\ref{tbl:binary_lj_params}. Pairs found within this shell but outside their pair-specific cutoff do not contribute to the force. The core radius and shell batch size are $r_c=1.20\sigma_{11}$ and $P=10$, respectively. All simulations use $\Delta t=0.01\tau_{\mathrm{LJ}}$ and $\bm\Gamma=0.5\bm I$. For Mi-RBL, $S_0=(\epsilon_{11}/\sigma_{11})^2=1$. Short pilot runs based on equilibrated kinetic statistics were used to select the tested ranges $0.25\le\alpha\le0.37$ and $0.91\le\beta\le0.99$. Each trajectory is equilibrated for $1.5\times10^5$ steps and followed by $4\times10^5$ production steps.

\begin{figure*}[htbp]
  \centering
  \includegraphics[width=0.95\textwidth]{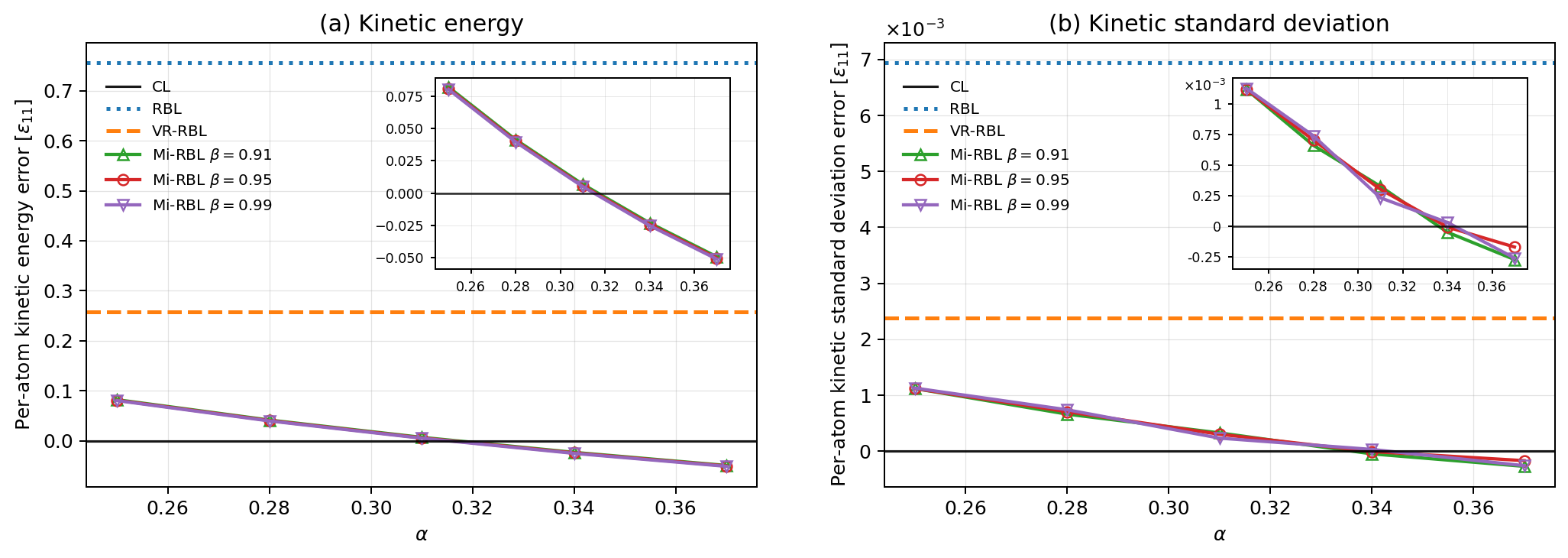}
  \caption[Binary LJ kinetic diagnostics]{Signed deviations of the per-atom kinetic energy from the CL reference for the Kob--Andersen mixture at $P=10$. (a) Kinetic energy error, $\Delta k=\bar{k}_{\mathrm{method}}-\bar{k}_{\mathrm{CL}}$; (b) error in the kinetic standard deviation, $\Delta\sigma_k=\sigma_{k,\mathrm{method}}-\sigma_{k,\mathrm{CL}}$. Mi-RBL is tested over $0.25\le\alpha\le0.37$ for $\beta=0.91$, $0.95$, and $0.99$. Positive and negative values indicate overestimation and underestimation of the corresponding CL. The insets magnify the Mi-RBL values near zero.}
  \label{fig:binary_kinetic_errors}
\end{figure*}

Figure~\ref{fig:binary_kinetic_errors}(a) shows that increasing $\alpha$ continuously shifts the Mi-RBL kinetic mean toward and then below the CL reference. At $\alpha=0.31$, $|\Delta k|$ remains below $6.82\times10^{-3}$ for all three values of $\beta$, compared with $2.58\times10^{-1}$ for VR-RBL and $7.56\times10^{-1}$ for RBL. For $\alpha\ge0.34$, the Mi-RBL error becomes negative, indicating that the kinetic mean falls below the CL value. This sign change can be understood from Eq.~\eqref{eq:temp_balance_general}. Increasing $\alpha$ first reduces the heating from the covariance term. At larger $\alpha$, the bias introduced by force scaling increasingly controls the kinetic mean. For each tested $\alpha$, the curves for $\beta=0.91$, $0.95$, and $0.99$ nearly overlap, indicating weak dependence on $\beta$.

Figure~\ref{fig:binary_kinetic_errors}(b) shows the error in $\sigma_k$, the standard deviation of the per-atom kinetic energy over time. RBL and VR-RBL both give positive $\Delta\sigma_k$, indicating larger kinetic fluctuations than CL. Across the tested Mi-RBL parameter values, $|\Delta\sigma_k|$ ranges from $6.56\times10^{-6}$ to $1.13\times10^{-3}$, compared with $2.38\times10^{-3}$ for VR-RBL and $6.94\times10^{-3}$ for RBL. Both kinetic errors are reduced at all tested $\alpha$ values.

\begin{figure*}[htbp]
  \centering
  \includegraphics[width=0.98\textwidth]{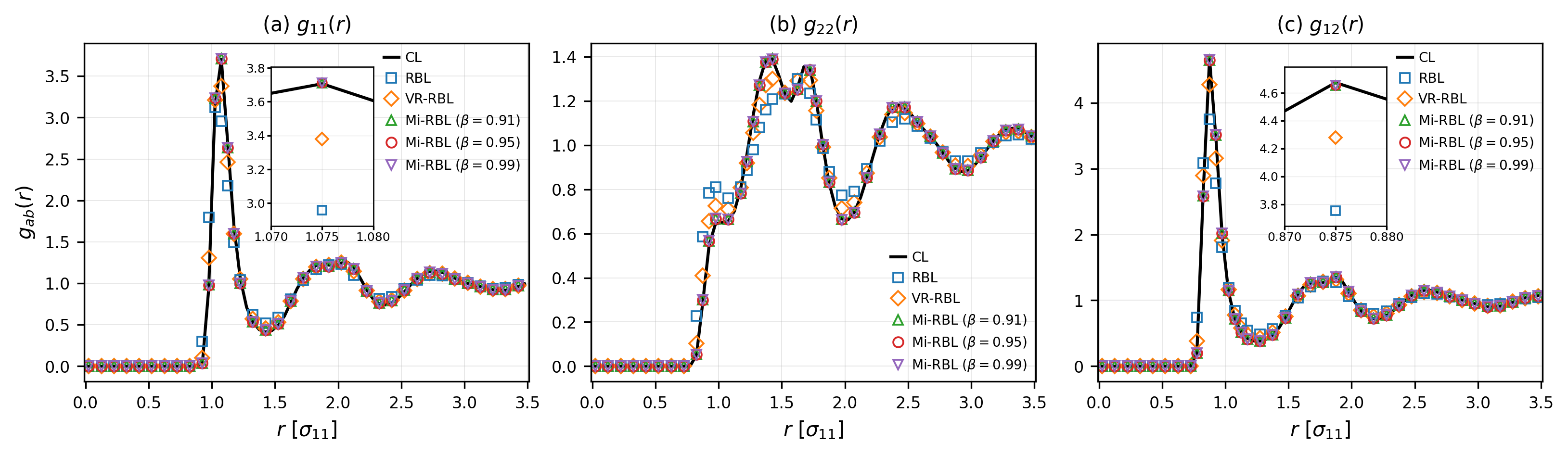}
  \caption[Binary LJ partial RDFs]{Partial radial distribution functions for the Kob--Andersen mixture at $P=10$: (a) $g_{11}(r)$, (b) $g_{22}(r)$, and (c) $g_{12}(r)$. The curves compare CL, RBL, VR-RBL, and Mi-RBL at $\alpha=0.31$ with $\beta=0.91$, $0.95$, and $0.99$.}
  \label{fig:binary_rdf}
\end{figure*}

Figure~\ref{fig:binary_rdf} compares the partial RDFs using the value $\alpha=0.31$ selected from the kinetic-mean error in Fig.~\ref{fig:binary_kinetic_errors}(a). RBL gives clear deviations from CL in all three partial RDFs. For $g_{11}(r)$ and $g_{12}(r)$, the main deviations appear near the first coordination shell, with lower first peaks. For $g_{22}(r)$, RBL does not reproduce the oscillatory structure. VR-RBL reduces these errors, but visible differences remain in all three pair types. The Mi-RBL curves are much closer to CL for all three pair types.

Table~\ref{tbl:binary_rdf_rmse} reports the corresponding root-mean-square errors (RMSEs). Mi-RBL has the smallest RMSEs for $g_{11}(r)$, $g_{12}(r)$, and $g_{22}(r)$. The VR-RBL errors are $3.1$--$8.5$ times larger than the Mi-RBL errors, and the RBL errors are $7.1$--$19.3$ times larger.

\begin{table}[htbp]
  \caption[Binary LJ RDF errors]{Root-mean-square errors of the three partial radial distribution functions relative to CL for the Kob--Andersen mixture at $P=10$. Mi-RBL uses $\alpha=0.31$ with the indicated $\beta$ values. Lower values indicate closer agreement with the CL pair structure.}
  \label{tbl:binary_rdf_rmse}
  \centering
  \begin{ruledtabular}
  \begin{tabular}{lccc}
    Method & $g_{11}(r)$ RMSE & $g_{12}(r)$ RMSE & $g_{22}(r)$ RMSE \\
    \hline
    RBL & $0.1615$ & $0.1708$ & $0.08517$ \\
    VR-RBL & $0.07072$ & $0.07573$ & $0.03980$ \\
    Mi-RBL ($\beta=0.91$) & $0.02282$ & $0.008870$ & $0.008640$ \\
    Mi-RBL ($\beta=0.95$) & $0.02258$ & $0.009129$ & $0.008470$ \\
    Mi-RBL ($\beta=0.99$) & $0.02276$ & $0.009122$ & $0.008898$ \\
  \end{tabular}
  \end{ruledtabular}
\end{table}
\subsection{Primitive Electrolyte}

The primitive electrolyte test applies the same rescaling idea to RBE and examines the cation-centered charge density profile at small batch sizes. We consider a bulk $3{:}1$ electrolyte. \cite{LIANG2022108332} The charge density profile characterizes the ionic screening around a tagged trivalent cation.

The system contains $800$ trivalent cations and $2400$ monovalent anions in a cubic periodic box of side length $L=25.6\sigma_{\mathrm{LJ}}$. The ions carry charges $q_+=+3$ and $q_-=-1$ and interact through Coulomb forces together with a repulsive shifted-truncated Lennard-Jones core cut off at $r_{\mathrm{LJ}}=2^{1/6}\sigma_{\mathrm{LJ}}$. The unit-charge Bjerrum length is $\ell_B=2.857\sigma_{\mathrm{LJ}}$. We use Lennard-Jones reduced units with $\sigma_{\mathrm{LJ}}=\epsilon_{\mathrm{LJ}}=m=1$ and $\tau_{\mathrm{LJ}}=\sigma_{\mathrm{LJ}}\sqrt{m/\epsilon_{\mathrm{LJ}}}$.

Unless otherwise stated, calculations use $T=1.0$, $\Delta t=0.0025\tau_{\mathrm{LJ}}$, and $\bm\Gamma=0.25\bm I$. The real-space Coulomb cutoff is $8.0\sigma_{\mathrm{LJ}}$, and PPPM with a target accuracy of $10^{-4}$ serves as the reference calculation. \cite{deserno_how_1998} RBE and Mi-RBE are tested at $P=5$, $10$, and $20$. Mi-RBE fixes $\beta=0.95$ and $S_0=(\epsilon_{\mathrm{LJ}}/\sigma_{\mathrm{LJ}})^2=1$. Short pilot runs were used to select $0.4\le\alpha\le0.6$ for $P=5$ and $10$, and $0.1\le\alpha\le0.3$ for $P=20$. 

To characterize the ionic atmosphere around a tagged cation, we use the cation-centered charge density profile. \cite{forsman_efficient_2024,kjellander2019dielectric,kjellander2020debye} The partial RDF $g_{ab}(r)$ gives the conditional number density $\rho_b g_{ab}(r)$ of species $b$ around species $a$. Weighting these conditional densities by the ionic charges gives
\begin{equation}
  \rho_q^+(r)
  =
  q_+\rho_+g_{++}(r)
  +
  q_-\rho_-g_{+-}(r),
  \label{eq:electrolyte_charge_density}
\end{equation}
where $\rho_+$ and $\rho_-$ are the bulk number densities of the two ionic species. A negative first-shell minimum reflects anion enrichment around the tagged cation.

\begin{figure*}[htbp]
  \centering
  \includegraphics[width=0.98\textwidth]{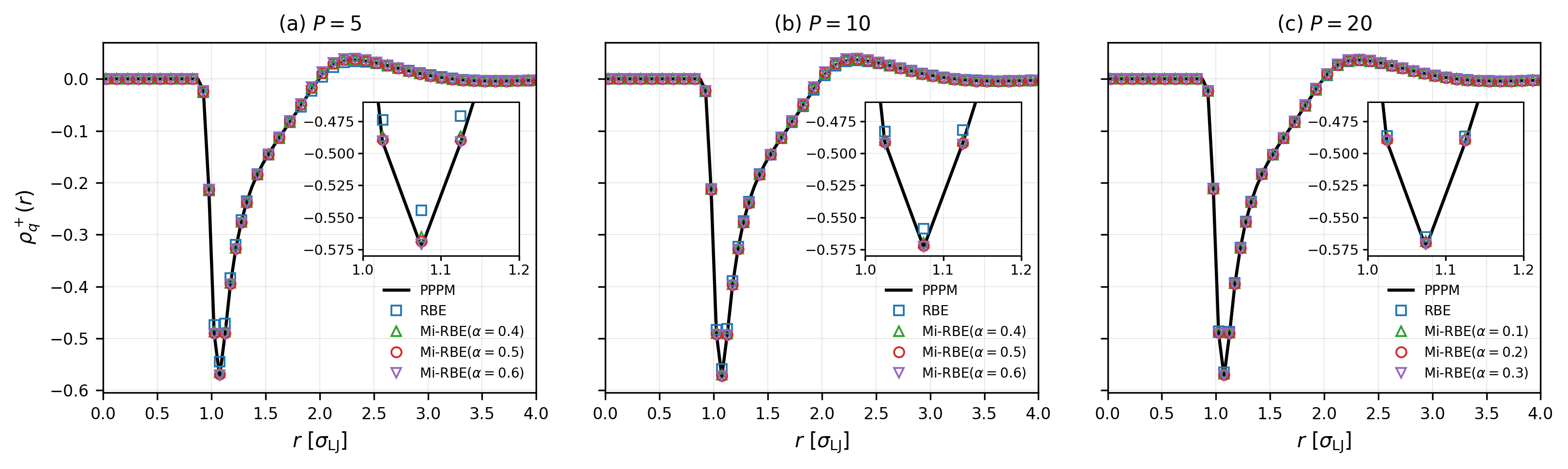}
  \caption[Primitive electrolyte charge density]{Cation-centered charge density $\rho_q^+(r)$ for the primitive electrolyte at (a) $P=5$, (b) $P=10$, and (c) $P=20$. The insets magnify the first-shell minimum.}
  \label{fig:electrolyte_charge_density}
\end{figure*}

\begin{table}[htbp]
  \caption{Root-mean-square errors of the cation-centered charge density relative to PPPM for the primitive electrolyte. Mi-RBE uses $\beta=0.95$, with the corresponding value of $\alpha$ listed in parentheses.}
  \label{tbl:electrolyte_charge_density_rmse}
  \centering
  \begin{ruledtabular}
  \begin{tabular}{ccl}
    
    Batch size $P$ & RBE RMSE & Mi-RBE RMSE ($\alpha$) \\
    \hline
    $5$  & $3.138\times10^{-3}$ & \begin{tabular}[t]{@{}l@{}}$9.889\times10^{-4}$ ($\alpha=0.4$)\\ $9.447\times10^{-4}$ ($\alpha=0.5$)\\ $1.136\times10^{-3}$ ($\alpha=0.6$)\end{tabular} \\
    $10$ & $1.541\times10^{-3}$ & \begin{tabular}[t]{@{}l@{}}$7.235\times10^{-4}$ ($\alpha=0.4$)\\ $9.009\times10^{-4}$ ($\alpha=0.5$)\\ $1.125\times10^{-3}$ ($\alpha=0.6$)\end{tabular} \\
    $20$ & $8.474\times10^{-4}$ & \begin{tabular}[t]{@{}l@{}}$5.214\times10^{-4}$ ($\alpha=0.1$)\\ $4.459\times10^{-4}$ ($\alpha=0.2$)\\ $4.993\times10^{-4}$ ($\alpha=0.3$)\end{tabular} \\
  \end{tabular}
  \end{ruledtabular}
\end{table}
Figure~\ref{fig:electrolyte_charge_density} compares the cation-centered charge density with the PPPM reference. At $P=5$ and $P=10$, RBE gives a first-shell minimum that is too shallow, indicating insufficient net anion enrichment around the cation. Mi-RBE comes closer to the depth of this minimum, even at $P=5$, and better reproduces the nearby PPPM profile. At $P=20$, RBE is already closer to PPPM, but the tested Mi-RBE parameters still give smaller deviations near the first coordination shell.

Table~\ref{tbl:electrolyte_charge_density_rmse} reports the RMSE values. At every batch size, each tested Mi-RBE value of $\alpha$ gives a smaller charge density RMSE than RBE. Across the tested $\alpha$ values, the RBE-to-Mi-RBE RMSE ratios are $2.8$--$3.3$, $1.4$--$2.1$, and $1.6$--$1.9$ for $P=5$, $10$, and $20$, respectively. Smaller $\alpha$ values remain effective at larger batch sizes, consistent with weaker random-batch fluctuations.

\begin{figure*}[t]
  \begin{minipage}[t]{0.48\textwidth}
    \centering
    \includegraphics[width=0.95\linewidth]{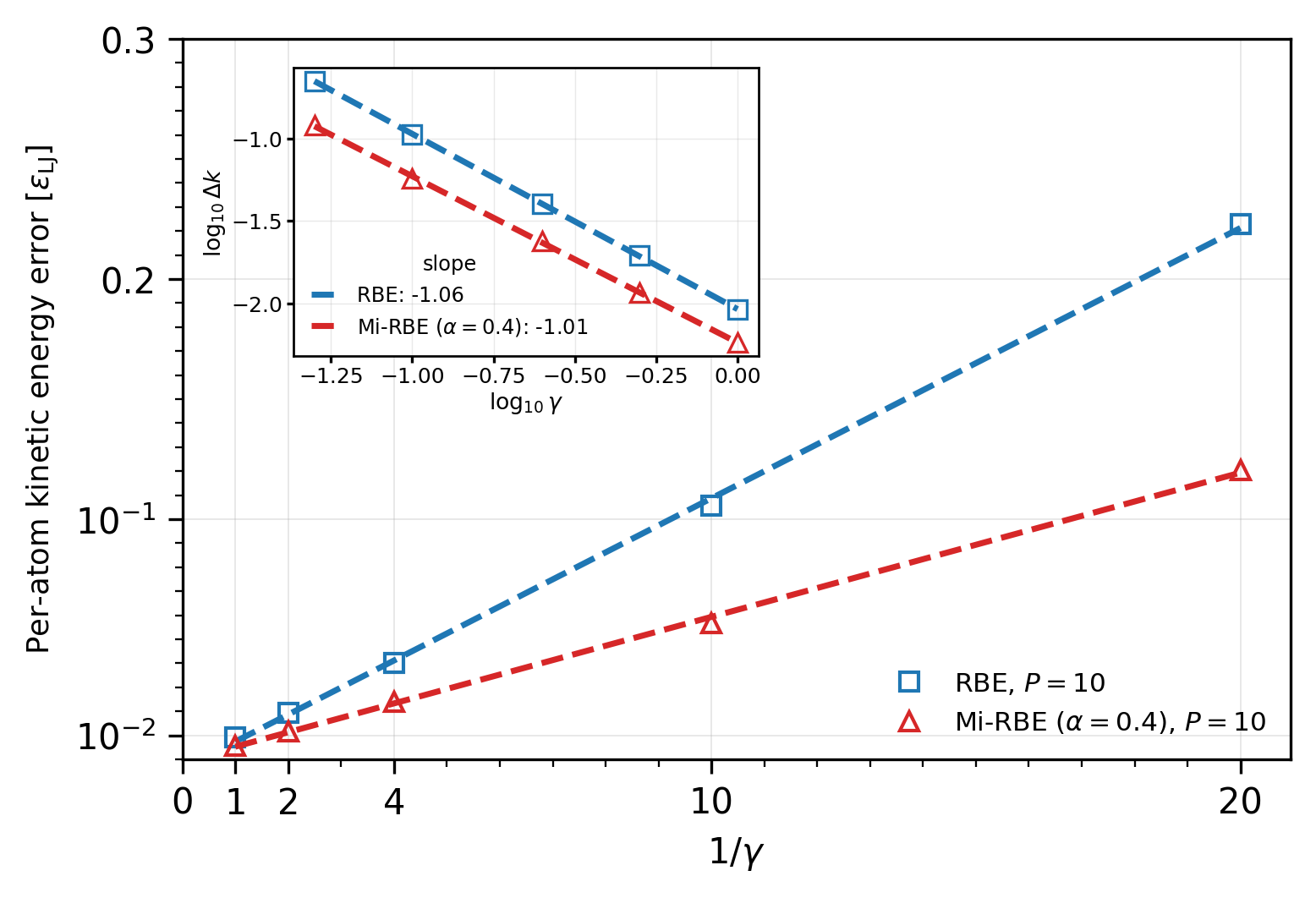}
    \caption[Primitive electrolyte kinetic error scaling with friction]{Per-atom kinetic energy error $\Delta k$ as a function of $1/\gamma$ for the primitive electrolyte at $P=10$ and $\Delta t=0.0025\tau_{\mathrm{LJ}}$, with $\bm\Gamma=\gamma\bm I$. Mi-RBE uses $(\alpha,\beta)=(0.4,0.95)$. The inset gives the slopes from linear fits of $\log_{10}\Delta k$ against $\log_{10}\gamma$.}
    \label{fig:electrolyte_kinetic_gamma}
  \end{minipage}\hfill
  \begin{minipage}[t]{0.48\textwidth}
    \centering
    \includegraphics[width=0.95\linewidth]{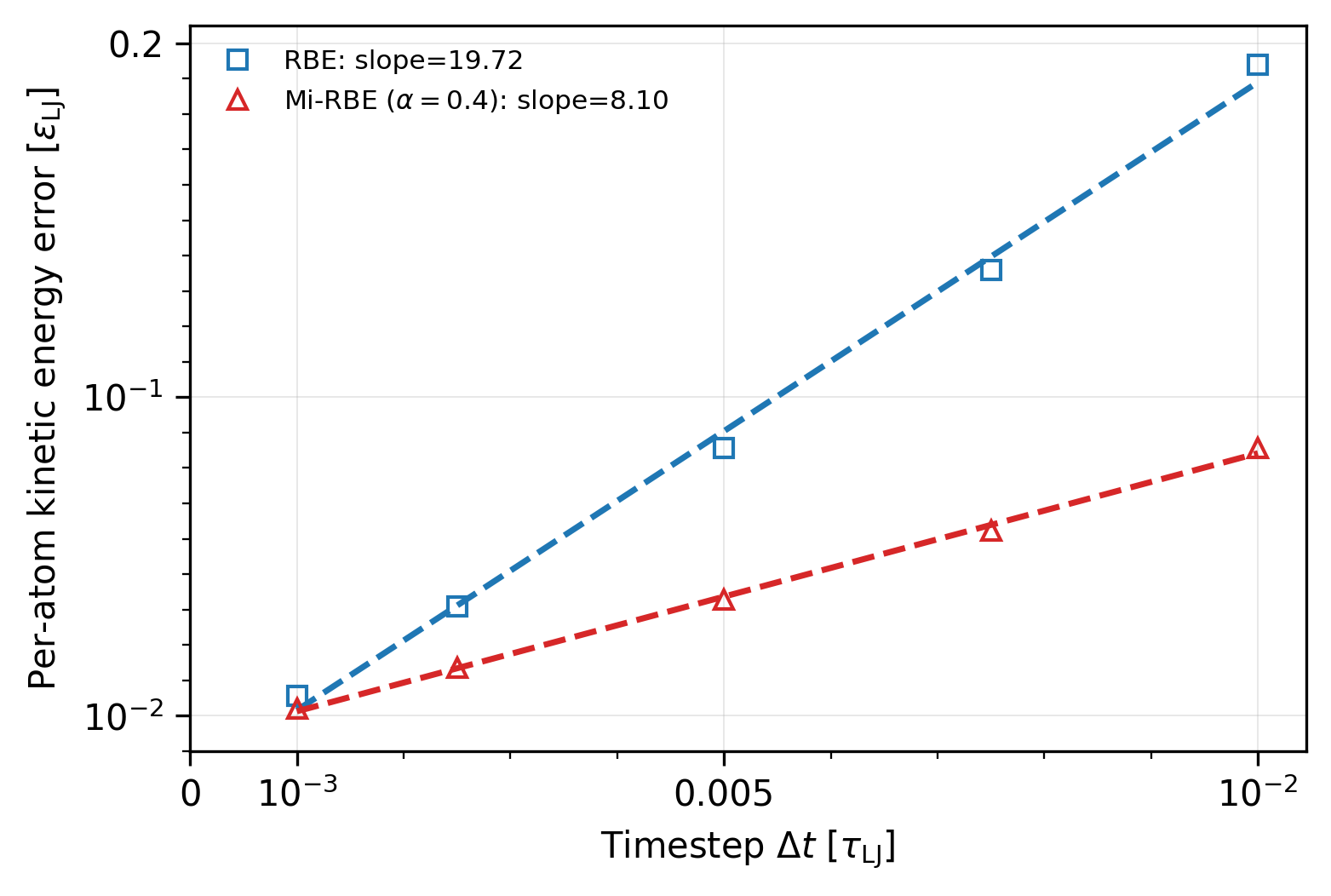}
    \caption[Primitive electrolyte kinetic error scaling with time step]{Per-atom kinetic energy error $\Delta k$ as a function of the time step for the primitive electrolyte at $P=10$ and $\bm\Gamma=0.25\bm I$. Mi-RBE uses $(\alpha,\beta)=(0.4,0.95)$. Linear fits give slopes of $19.72$ for RBE and $8.10$ for Mi-RBE.}
    \label{fig:electrolyte_kinetic_timestep}
  \end{minipage}
\end{figure*}

We use Eq.~\eqref{eq:temp_balance_general} for two kinetic error tests in the primitive electrolyte. We define the per-atom kinetic energy error as $\Delta k=\bar{k}-k_T$, where $k_T=3k_BT/2$, and set $T=1.0$. With $(\alpha,\beta)=(0.4,0.95)$ fixed for Mi-RBE, the first test varies $1/\gamma$ at fixed $\Delta t$, and the second varies $\Delta t$ at fixed $\gamma$.

Figure~\ref{fig:electrolyte_kinetic_gamma} tests the friction dependence at $P=10$ and $\Delta t=0.0025\tau_{\mathrm{LJ}}$. Since the kinetic energy errors are positive over this range, we take their logarithms directly. The log--log fits give exponents of $-1.06$ for RBE and $-1.01$ for Mi-RBE, both close to the predicted value of $-1$, indicating that the error is approximately $O(1/\gamma)$ as in Eq.~\eqref{eq:temp_balance_general}. At the same $\gamma$, Mi-RBE gives smaller kinetic energy errors than RBE.

At fixed $P=10$ and $\bm\Gamma=0.25\bm I$, Figure~\ref{fig:electrolyte_kinetic_timestep} shows an approximately linear increase in $\Delta k$ with the time step. The fitted slopes are $19.72$ for RBE and $8.10$ for Mi-RBE. The Mi-RBE slope is therefore less than half the RBE slope, showing that rescaling reduces the growth rate of the kinetic error with the time step by more than half. This is consistent with Eq.~\eqref{eq:temp_balance_general}: after rescaling, the covariance term contributes less to the kinetic heating. These tests assess the combined kinetic effect of force rescaling without separately resolving the bias and covariance contributions.

\begin{figure}[!t]
  \centering
  \includegraphics[width=0.95\columnwidth]{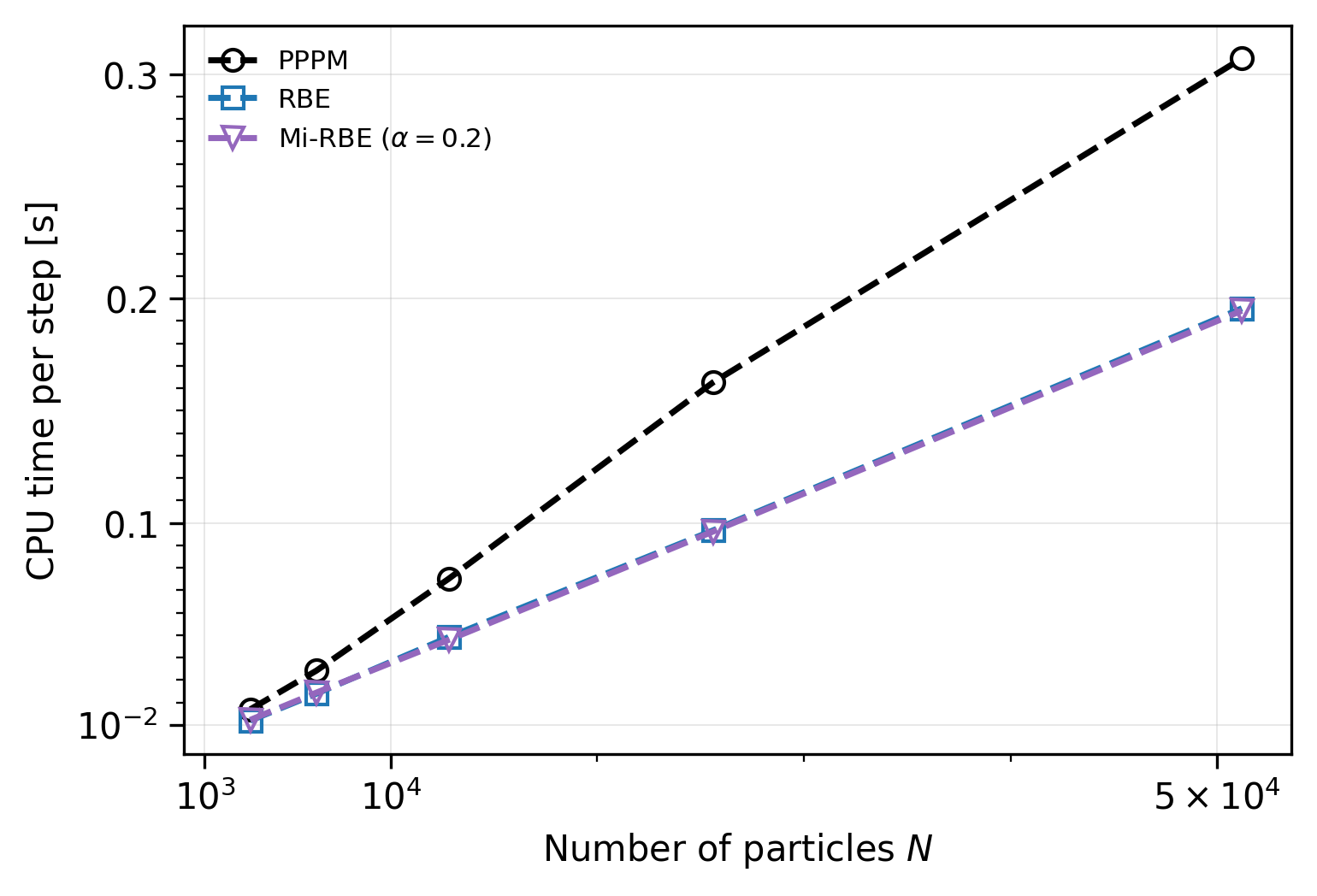}
  \caption[Electrolyte CPU time per step]{CPU time per integration step versus particle number $N$ for RBE, Mi-RBE, and PPPM in the primitive electrolyte. RBE and Mi-RBE use $P=20$; Mi-RBE uses $(\alpha,\beta)=(0.2,0.95)$, and PPPM uses a target accuracy of $10^{-4}$.}
  \label{fig:electrolyte_timing}
\end{figure}

Finally, we examine the additional cost introduced by the particlewise rescaling step. At fixed $P=20$, Figure~\ref{fig:electrolyte_timing} shows that RBE and Mi-RBE have very similar CPU times per step as $N$ increases, so the lagged moment update and scalar gain add little extra cost. The CPU times of both random-batch methods grow nearly linearly with $N$, in agreement with the $O(N)$ cost at fixed $P$, and stay below PPPM. Thus, Mi-RBE improves the charge density and kinetic results at little extra cost.

\section{Conclusion}

We have introduced moment-informed particlewise force rescaling for random-batch Langevin dynamics. The construction uses a lagged estimate of each particle's random-batch force intensity to rescale its magnitude while preserving its direction. The formal finite-time estimate separates the force-scaling contribution from the transformed random-batch covariance. The steady-state kinetic-temperature approximation then connects these terms to the kinetic error.

In the numerical tests, Mi-RBL gives smaller kinetic errors and better MSD and RDF results than RBL and VR-RBL. For the electrolyte, Mi-RBE better reproduces the first-shell anion enrichment, and it reduces the RBE kinetic error at little extra cost. Within the tested ranges, $\alpha$ controls the balance between the transformed random-batch covariance and the force-scaling contribution, whereas the results depend only weakly on $\beta$ for the tested $\beta$ values in the binary-mixture test. In both the RBL and RBE tests, the effective rescaling strength decreases with increasing batch size, consistent with the reduction in random-batch fluctuations. These systematic trends provide practical guidance for parameter selection.

\begin{acknowledgments}
This work is supported by the National Natural Science Foundation of China (Grants No. 12426304 and
12325113). The authors would like to thank the support from the SJTU Kunpeng \& Ascend
Center of Excellence.
\end{acknowledgments}
\section*{Author Declarations}

\subsection*{Conflict of Interest}

The authors have no conflicts to disclose.

\subsection*{Author Contributions}

Xingguo Wu: Conceptualization, Methodology, Software, Investigation, Validation, Visualization, Writing (original draft).
Yangshuai Wang: Conceptualization, Methodology, Supervision, Formal analysis, Investigation, Writing (review and editing).
Zhenli Xu: Conceptualization, Methodology, Supervision, Project administration, Funding acquisition, Writing (review and editing).

\section*{Data Availability Statement}

The data that support the findings of this study are available from the corresponding authors upon reasonable request.

\appendix

\section{Derivation of the Formal Finite-Time Estimate}

Here we separate the error due to force rescaling and random batching from the discretization error of the velocity-Verlet integrator. We compare the exact Langevin dynamics with an auxiliary dynamics with rescaled random-batch forces. In this auxiliary dynamics, a new random batch is chosen at $t_n=n\Delta t$. The corresponding random-batch force error is kept fixed on $[t_n,t_{n+1})$, while the exact force and the Langevin thermostat are treated in continuous time. Thus, the estimate below does not include the integrator error used in the numerical simulations.

Fix an initial phase-space state $R=(\bm r_0,\bm v_0)$. The exact dynamics and the auxiliary dynamics start from the same state $R$ and are driven by the same $3N$-dimensional Wiener process. Let $\mathcal F_n$ denote the information before the step-$n$ batch is selected. It is generated by the initial state $R$, the Wiener process up to time $t_n$, and all random batches chosen through step $n-1$. Thus, $\mathcal F_n$ contains the states of both dynamics at the beginning of step $n$ and the lagged EMA states $\{s_{i,n}\}_{i=1}^N$, but not the random batch selected at step $n$.

Write
\[
  \widehat{\bm r}_n=\widehat{\bm r}(t_n),
  \qquad
  \widehat{\bm v}_n=\widehat{\bm v}(t_n).
\]
Because the moment-informed gain is determined from the lagged state, the matrix
\begin{equation}
  \bm G_n(\alpha)
  =
  \operatorname{diag}
  (g_{1,n}\bm I_3,\ldots,g_{N,n}\bm I_3)
\end{equation}
is $\mathcal F_n$-measurable; initially, $\bm G_0=\bm I$. Conditional on $\mathcal F_n$, the current random-batch estimator satisfies
\[
\begin{aligned}
  \widetilde{\bm F}_{b,n}
  &=
  \bm F_b(\widehat{\bm r}_n)+\bm\zeta_n, \\
  \E[\bm\zeta_n\mid\mathcal F_n]
  &=\bm0,
  \quad
  \Cov(\bm\zeta_n\mid\mathcal F_n)
  =\bm\Sigma_n.
\end{aligned}
\]
The force error after projection can be written as
\begin{equation}
\begin{aligned}
  \bm e_n
  &=\bm b_n+\bm\xi_n, \\
  \bm b_n
  &=
  \mathsf P_0(\bm G_n-\bm I)\bm F_b(\widehat{\bm r}_n), \\
  \bm\xi_n
  &=
  \mathsf P_0\bm G_n\bm\zeta_n.
\end{aligned}
\end{equation}
Thus, $\bm b_n$ is $\mathcal F_n$-measurable, and the lagged construction gives $\E[\bm\xi_n\mid\mathcal F_n]=\bm0$.

Define the mass matrix by
\[
  \bm M
  =
  \operatorname{diag}
  (m_1\bm I_3,\ldots,m_N\bm I_3).
\]
On each interval $[t_n,t_{n+1})$, the synchronous coupling is
\begin{equation}
\left\{
\begin{aligned}
\dd\bm r
&= \bm v\,\dd t, \\
\bm M\,\dd\bm v
&=
\left[\bm F(\bm r)-\gamma\bm M\bm v\right]\dd t \\
&\quad+
\sqrt{2\gamma k_B T}\,
\bm M^{1/2}\,\dd\bm W .
\end{aligned}
\right.
\label{eq:exact_dynamics}
\end{equation}

\begin{equation}
\left\{
\begin{aligned}
\dd\widehat{\bm r}
&= \widehat{\bm v}\,\dd t, \\
\bm M\,\dd\widehat{\bm v}
&=
\left[
\bm F(\widehat{\bm r})
+\bm b_n+\bm\xi_n
-\gamma\bm M\widehat{\bm v}
\right]\dd t \\
&\quad+
\sqrt{2\gamma k_B T}\,
\bm M^{1/2}\,\dd\bm W .
\end{aligned}
\right.
\label{eq:modified_dynamics}
\end{equation}
Here $\bm F=\bm F_a+\bm F_b$ denotes the exact force from particle interactions. In the auxiliary dynamics, $\bm b_n$ and $\bm\xi_n$ are evaluated at $\widehat{\bm r}_n$ and then kept fixed on $[t_n,t_{n+1})$. 

Define
\begin{equation}
  \delta\bm r=\bm r-\widehat{\bm r},
  \qquad
  \delta\bm v=\bm v-\widehat{\bm v}.
\end{equation}
Since the two dynamics start from the same state, $\delta\bm r(0)=\delta\bm v(0)=\bm0$.
Subtracting the two systems yields
\begin{equation}
\left\{
\begin{aligned}
\dd\delta\bm r
&= \delta\bm v\,\dd t, \\[3pt]
\bm M\,\dd\delta\bm v
&=
\left[
\bm F(\bm r)
-\bm F(\widehat{\bm r})
-\bm b_n
-\bm\xi_n
-\gamma\bm M\delta\bm v
\right]\dd t .
\end{aligned}
\right.
\label{eq:local_error_sde}
\end{equation}
The Brownian term does not appear in Eq.~\eqref{eq:local_error_sde} because the same Wiener process drives both dynamics. For a fixed initial phase-space state $R$, define
\begin{equation}
\begin{aligned}
  E_N^R(t)
  &:={}
  \frac{1}{N}\E_R\left[
    \|\delta\bm r(t)\|^2
    +
    \|\delta\bm v(t)\|^2
  \right],
  \qquad 0\le t\le t^*, \\
  E_N^R(0)&=0.
\end{aligned}
\end{equation}
Differentiating $E_N^R(t)$ gives
\begin{equation}
\begin{aligned}
  \frac{\dd E_N^R}{\dd t}
  ={}&
  \frac{2}{N}\E_R[\delta\bm r\cdot\delta\bm v] \\
  +&\frac{2}{N}\E_R\left[\delta\bm v\cdot
  \bm M^{-1}(\bm F(\bm r)-\bm F(\widehat{\bm r})-\bm b_n-\bm\xi_n)
  \right]\\
  &-\frac{2\gamma}{N}\E_R\|\delta\bm v\|^2 .
\end{aligned}
\end{equation}
Using the uniform bounds on the particle masses and the Lipschitz continuity of the force, we obtain
\begin{equation}
\begin{aligned}
  \frac{\dd E_N^R}{\dd t}
  \le{}&
  C E_N^R(t)
  +\frac{C}{N}\E_R\|\bm b_n\|^2
  +\frac{C}{N}\left|\E_R[\delta\bm v(t)\cdot\bm\xi_n]\right| .
\end{aligned}
\label{eq:error_differential_ineq}
\end{equation}
Since $\delta\bm v(t_n)$ is $\mathcal F_n$-measurable,
\begin{equation}
\begin{aligned}
  \E_R[\delta\bm v(t_n)\cdot\bm\xi_n]
  &=
  \E_R\left[
    \delta\bm v(t_n)\cdot
    \E(\bm\xi_n\mid\mathcal F_n)
  \right] \\
  &=0.
\end{aligned}
\end{equation}
Consequently, for $t\in[t_n,t_{n+1})$,
\[
  \E_R[\delta\bm v(t)\cdot\bm\xi_n]
  =
  \E_R\left[
    (\delta\bm v(t)-\delta\bm v(t_n))
    \cdot\bm\xi_n
  \right].
\]
The integral form of Eq.~\eqref{eq:local_error_sde}, together with the mass bounds, the Lipschitz continuity of $\bm F$, and the Cauchy--Schwarz and Young inequalities, gives
\begin{equation}
\begin{aligned}
  &\frac{1}{N}\left|\E_R[\delta\bm v(t)\cdot\bm\xi_n]\right| \\
  &\qquad\le
  C(t-t_n)\sup_{s\in[t_n,t]}E_N^R(s) \\
  &\qquad\quad
  +\frac{C(t-t_n)}{N}\E_R\|\bm b_n\|^2 \\
  &\qquad\quad
  +\frac{C(t-t_n)}{N}\E_R\|\bm\xi_n\|^2 .
\end{aligned}
\end{equation}
Substituting the bound on $\E_R[\delta\bm v(t)\cdot\bm\xi_n]$ into Eq.~\eqref{eq:error_differential_ineq} and integrating over $[t_n,t_{n+1}]$ gives, for sufficiently small $\Delta t$,
\begin{equation}
\begin{aligned}
  E_N^R(t_{n+1})
  \le{}&
  (1+C\Delta t)E_N^R(t_n)
  +\frac{C\Delta t}{N}\E_R\|\bm b_n\|^2 \\
  &+\frac{C\Delta t^2}{N}\E_R\|\bm\xi_n\|^2 .
\end{aligned}
\end{equation}
Using the conditional covariance of $\bm\xi_n$, define
\begin{equation}
\begin{aligned}
  \bm Q_n(\alpha)
  &:={}
  \Cov(\bm\xi_n\mid\mathcal F_n)
  \\
  &=
  \mathsf P_0
  \bm G_n\bm\Sigma_n\bm G_n^T
  \mathsf P_0^T, \\
  q_n(\alpha)
  &:={}
  \operatorname{Tr}\bm Q_n(\alpha).
\end{aligned}
\end{equation}
Since $\bm G_n$ is $\mathcal F_n$-measurable,
\begin{equation}
  \E(\bm\xi_n\mid\mathcal F_n)=0,
  \qquad
  \E(\|\bm\xi_n\|^2\mid\mathcal F_n)=q_n(\alpha),
\end{equation}
and hence
\begin{equation}
  E_N^R(t_{n+1})
  \le
  (1+C\Delta t)E_N^R(t_n)
  +\frac{C\Delta t}{N}\E_R\|\bm b_n\|^2
  +\frac{C\Delta t^2}{N}\E_R q_n(\alpha) .
  \label{eq:local_recursion}
\end{equation}

Let $K:=N_{t^*}=t^*/\Delta t\in\mathbb N$. We define the following time averages of the bias and covariance terms:
\begin{equation}
\begin{aligned}
\mathcal B_{t^*,N}(\alpha)
&:=
\sup_R \frac{1}{Nt^*}
\sum_{n=0}^{K-1}
\Delta t\,\E_R\|\bm b_n\|^2, \\
\mathcal Q_{t^*,N}(\alpha)
&:=
\sup_R \frac{1}{Nt^*}
\sum_{n=0}^{K-1}
\Delta t\,\E_R q_n(\alpha).
\end{aligned}
\label{eq:accumulated_error_terms_appendix}
\end{equation}
When $\alpha$ is small, Eq.~\eqref{eq:bias_expansion} gives a uniform bound under the boundedness assumptions on the force and on the logarithmic factors in $\bm L_n$. Specifically, there exist $\alpha_0>0$ and $C_b<\infty$, independent of $N$, $n$, $R$, and $\Delta t$, such that
\begin{equation}
\sup_R\max_{0\le n<K}
\frac{1}{N}\E_R\|\bm b_n\|^2
\le C_b\alpha^2,
\qquad
0\le\alpha\le\alpha_0.
\label{eq:uniform_bias_bound}
\end{equation}
Consequently, $\mathcal B_{t^*,N}(\alpha)\le C_b\alpha^2$.

Starting from $E_N^R(0)=0$, we iterate Eq.~\eqref{eq:local_recursion} and apply the discrete Gronwall inequality. For fixed $t^*$ and sufficiently small $\Delta t$, this gives
\begin{equation}
E_N^R(t^*)
\le
C(t^*)
\left[
\mathcal B_{t^*,N}(\alpha)
+
\Delta t\,\mathcal Q_{t^*,N}(\alpha)
\right].
\label{eq:tradeoff_appendix}
\end{equation}
Here $C(t^*)$ is independent of $N$, $\alpha$, and $\Delta t$.

Let $\mu_{t^*}^R$ and $\widehat\mu_{t^*}^{\alpha,R}$ denote the phase-space distributions of the exact and auxiliary dynamics at time $t^*$, respectively. The synchronous coupling gives a joint distribution $\lambda_R^{\mathrm{sync}}$ such that
\[
\lambda_R^{\mathrm{sync}}
\in
\Pi\left(
\mu_{t^*}^R,
\widehat\mu_{t^*}^{\alpha,R}
\right).
\]
By the definition of $W_{2,N}$,
\begin{equation}
\begin{aligned}
W_{2,N}^2\left(
\mu_{t^*}^R,
\widehat\mu_{t^*}^{\alpha,R}
\right)
&\le
\int
\frac{1}{N}\|\mathbf x-\mathbf y\|_2^2
\,\dd\lambda_R^{\mathrm{sync}}(\mathbf x,\mathbf y) \\
&=E_N^R(t^*) \\
&\le C(t^*)
\left[
\mathcal B_{t^*,N}(\alpha)
+
\Delta t\,\mathcal Q_{t^*,N}(\alpha)
\right].
\end{aligned}
\label{eq:wasserstein_closure}
\end{equation}
Taking square roots and then the supremum over $R$ yields Eq.~\eqref{eq:tradeoff}.

\section{Approximate steady-state kinetic-temperature error estimate and the VR-RBL Covariance Correction}

\subsection{Approximate kinetic-temperature error estimate}

To estimate $T_\alpha-T$, we use a continuous time approximation. We take equal particle masses $m$ and regard the lagged gain and $\Cov(\bm\xi_\alpha\mid\mathcal F_n)=\bm Q_\alpha$ as fixed during a time step. In the discrete velocity update, the random-batch force error enters as $\Delta t\,\bm\xi_\alpha$, whose covariance is $\Delta t^2\bm Q_\alpha$. We replace this contribution by an independent Brownian term with the same covariance over the interval $\Delta t$. Once the system has reached a steady state, the averaged kinetic and potential energies are time independent.

At stationarity, each step is treated in the same way, so we consider one time step and omit the subscript $n$ below.
\begin{equation}
  \bm b_\alpha
  =
  \mathsf P_0(\bm G_\alpha-\bm I)\bm F_b(\bm r),
  \qquad
  \bm\xi_\alpha
  =
  \mathsf P_0\bm G_\alpha\bm\zeta,
\end{equation}
where $\E(\bm\zeta\mid\mathcal F_n)=\bm0$ and $\Cov(\bm\zeta\mid\mathcal F_n)=\bm\Sigma$. Because $\bm G_\alpha$ is $\mathcal F_n$-measurable, the transformed covariance is
\begin{equation}
  \bm Q_\alpha
  =
  \Cov(\bm\xi_\alpha\mid\mathcal F_n)
  =
  \mathsf P_0
  \bm G_\alpha\bm\Sigma\bm G_\alpha^T
  \mathsf P_0^T.
\end{equation}
The equivalent Brownian term is defined by equating its covariance over one time step with that of $\Delta t\,\bm\xi_\alpha$:
\begin{equation}
  2\bm D_\alpha\Delta t
  =
  \Delta t^2\bm Q_\alpha,
  \qquad
  \bm D_\alpha
  =
  \frac{\Delta t}{2}\bm Q_\alpha.
\end{equation}
The corresponding effective Langevin equation is
\begin{equation}
\begin{aligned}
  m\,\dd\bm v
  ={}&
  [\bm F(\bm r)+\bm b_\alpha-\gamma m\bm v]\,\dd t \\
  &+ \sqrt{2\gamma m k_B T}\,\dd\bm W
  + (2\bm D_\alpha)^{1/2}\,\dd\bm B,
\end{aligned}
\label{eq:effective_langevin_balance}
\end{equation}
where $\bm W$ and $\bm B$ are independent standard $3N$-dimensional Wiener processes. We write $\langle A\rangle=N^{-1}\E_{\mathrm{ss}}[A]$ for stationary averages per particle. Applying It\^o's formula to Eq.~\eqref{eq:effective_langevin_balance} gives
\begin{equation}
\begin{aligned}
  \frac{\dd}{\dd t}\langle \|\bm v\|^2\rangle
  ={}&
  \frac{2}{m}\langle \bm v\cdot\bm F(\bm r)\rangle
  +
  \frac{2}{m}\langle \bm v\cdot\bm b_\alpha\rangle
  -
  2\gamma\langle \|\bm v\|^2\rangle \\
  &+
  \frac{6\gamma k_B T}{m}
  +
  \frac{\Delta t}{m^2}
  \left\langle\operatorname{Tr}\bm Q_\alpha\right\rangle.
\end{aligned}
\end{equation}
Let $U(\bm r)$ be the total potential energy, so that $\bm F(\bm r)=-\nabla U(\bm r)$. At stationarity,
\begin{equation}
  \frac{\dd}{\dd t}\langle U(\bm r)\rangle
  =
  -\langle \bm v\cdot\bm F(\bm r)\rangle
  =0.
  \label{eq:stationary_potential_balance}
\end{equation}
The kinetic average is also stationary:
\begin{equation}
  \frac{\dd}{\dd t}\langle\|\bm v\|^2\rangle=0.
  \label{eq:stationary_kinetic_balance}
\end{equation}
For a given rescaling strength $\alpha$, define
\begin{equation}
  T_\alpha
  =
  \frac{m\langle\|\bm v\|^2\rangle}{3k_B}
\end{equation}
Using Eqs.~\eqref{eq:stationary_potential_balance} and \eqref{eq:stationary_kinetic_balance} gives
\begin{equation}
  T_\alpha-T
  \approx
  \frac{1}{3\gamma k_B}
  \langle \bm v\cdot\bm b_\alpha\rangle
  +
  \frac{\Delta t}{6\gamma m k_B}
  \left\langle\operatorname{Tr}\bm Q_\alpha\right\rangle,
  \label{eq:temp_balance_appendix}
\end{equation}
which is Eq.~\eqref{eq:temp_balance_general}. The force-scaling term can have either sign. The covariance term is nonnegative because $\bm Q_\alpha$ is positive semidefinite. 

\subsection{Particlewise covariance correction in VR-RBL}

Using the same notation, $\bm\zeta_n$ is the raw RBL force error before applying $\mathsf P_0$, and $\bm\Sigma_n=\Cov(\bm\zeta_n\mid\mathcal F_n)$. The covariance after projection is
\begin{equation}
\begin{aligned}
\bm Q_{0,n}
&:=
\Cov(\mathsf P_0\bm\zeta_n\mid\mathcal F_n) \\
&=
\mathsf P_0\bm\Sigma_n\mathsf P_0^T .
\end{aligned}
\label{eq:vrrbl_projected_covariance}
\end{equation}
For the RBL tests, VR-RBL estimates the covariance of each particle's force error,
\[
\bm\Sigma_{i,n}
:=
\Cov(\bm\zeta_{i,n}\mid\mathcal F_n)
\]
before subtracting the average force over all particles.

Because shell neighbors are sampled without replacement in the VR-RBL comparison, the covariance estimate uses a finite-population correction:
\begin{equation}
\begin{aligned}
\widehat{\bm\Sigma}_{i,n}
={}&
\frac{N_{i,n}^s(N_{i,n}^s-P)}
     {P(P-1)}
\sum_{j\in C_{i,n}}
\left(
\bm f_{ij}-\overline{\bm f}_{i,n}
\right) \\
&\qquad\qquad\times
\left(
\bm f_{ij}-\overline{\bm f}_{i,n}
\right)^T, \\
\overline{\bm f}_{i,n}
={}&
\frac{1}{P}
\sum_{j\in C_{i,n}}\bm f_{ij}.
\end{aligned}
\label{eq:vrrbl_finite_population}
\end{equation}
When $N_{i,n}^s\le P$, all shell neighbors are used and the shell-force estimator is exact, so we set $\widehat{\bm\Sigma}_{i,n}=\bm0$. This estimator is unbiased:
\begin{equation}
\E\left(
\widehat{\bm\Sigma}_{i,n}
\mid\mathcal F_n
\right)
=
\bm\Sigma_{i,n}.
\label{eq:vrrbl_covariance_unbiased}
\end{equation}
This form differs from Ref.~\citenum{xu2024vrrbl}, where the finite-population correction is not used. For the covariance calculation below, write
\begin{equation}
\widehat{\bm\Sigma}^{\mathrm{pw}}_n
=
\operatorname{diag}
\left(
\widehat{\bm\Sigma}_{1,n},
\ldots,
\widehat{\bm\Sigma}_{N,n}
\right).
\end{equation}

For particle $i$, the standard Brownian increment from the Langevin thermostat has covariance
\begin{equation}
\bm C_{i,n}^{\mathrm{th}}
=
2\gamma m_i k_B T\Delta t\,\bm I_3.
\end{equation}
For particle $i$, VR-RBL subtracts the estimate in Eq.~\eqref{eq:vrrbl_finite_population} from the covariance of the Brownian increment:
\begin{equation}
\widetilde{\bm C}_{i,n}^{\mathrm{VR}}
=
\bm C_{i,n}^{\mathrm{th}}
-
\Delta t^2\widehat{\bm\Sigma}_{i,n}.
\end{equation}
If $\widetilde{\bm C}_{i,n}^{\mathrm{VR}}$ is not positive semidefinite, it cannot be used as a covariance matrix. The Brownian covariance used in the simulations is therefore
\begin{equation}
\bm C_{i,n}^{\mathrm{VR}}
=
\begin{cases}
\widetilde{\bm C}_{i,n}^{\mathrm{VR}},
&
\lambda_{\min}
\left(
\widetilde{\bm C}_{i,n}^{\mathrm{VR}}
\right)
\ge0, \\[3pt]
\bm0,
&
\lambda_{\min}
\left(
\widetilde{\bm C}_{i,n}^{\mathrm{VR}}
\right)
<0.
\end{cases}
\label{eq:vrrbl_fallback}
\end{equation}
Here $\lambda_{\min}(\cdot)$ denotes the smallest eigenvalue. Conditional on the sampled batch, the Brownian increment has zero mean and covariance $\bm C_{i,n}^{\mathrm{VR}}$; the second branch sets this increment to zero for particle $i$ at step $n$.

Let $\bm C_n^{\mathrm{th}}$ and $\bm C_n^{\mathrm{VR}}$ have diagonal particle blocks $\bm C_{i,n}^{\mathrm{th}}$ and $\bm C_{i,n}^{\mathrm{VR}}$. Conditional on $\mathcal F_n$, the covariance of the random contribution over this step is
\begin{equation}
\begin{aligned}
\bm C_n^{\mathrm{tot}}
&=
\Delta t^2\bm Q_{0,n}
+
\E\left(
\bm C_n^{\mathrm{VR}}
\mid\mathcal F_n
\right) \\
&=
\bm C_n^{\mathrm{th}}
+
\Delta t^2\bm R_n^{\mathrm{VR}},
\end{aligned}
\end{equation}
where the residual covariance is
\begin{equation}
\bm R_n^{\mathrm{VR}}
:=
\bm Q_{0,n}
+
\Delta t^{-2}
\left[
\E\left(
\bm C_n^{\mathrm{VR}}
\mid\mathcal F_n
\right)
-
\bm C_n^{\mathrm{th}}
\right].
\label{eq:vrrbl_residual_covariance}
\end{equation}
The residual $\bm R_n^{\mathrm{VR}}$ contains the effects of the projection by $\mathsf P_0$ and the fallback in Eq.~\eqref{eq:vrrbl_fallback}.

Assuming the same particle mass $m$, define
\[
\left\langle
\operatorname{Tr}\bm R^{\mathrm{VR}}
\right\rangle
:=
\frac{1}{N}
\E_{\mathrm{ss}}
\operatorname{Tr}\bm R_n^{\mathrm{VR}}.
\]
The kinetic-temperature estimate then gives the mean temperature shift
\begin{equation}
T_{\VRRBL}-T
\approx
\frac{\Delta t}{6\gamma m k_B}
\left\langle
\operatorname{Tr}\bm R^{\mathrm{VR}}
\right\rangle.
\label{eq:vrrbl_temperature_shift}
\end{equation}

If the fallback is inactive for all particles and sampled batches, Eq.~\eqref{eq:vrrbl_residual_covariance} reduces to
\begin{equation}
\bm R_n^{\mathrm{VR}}
=
\bm Q_{0,n}
-
\E\left(
\widehat{\bm\Sigma}^{\mathrm{pw}}_n
\mid\mathcal F_n
\right).
\label{eq:vrrbl_residual_no_fallback}
\end{equation}
Thus, $\widehat{\bm\Sigma}^{\mathrm{pw}}_n$ does not necessarily cancel $\bm Q_{0,n}$. The projection by $\mathsf P_0$ and possible covariances between different particles are still present.

Eq.~\eqref{eq:vrrbl_temperature_shift} provides a simple explanation for the mean temperature drift caused by the VR-RBL correction, while a rigorous quantitative analysis of the long-time equilibrium distribution is beyond the scope of this estimate.
\bibliography{mi_rbl_refs}

%merlin.mbs aipnum4-1.bst 2010-07-25 4.21a (PWD, AO, DPC) hacked
%Control: key (0)
%Control: author (8) initials jnrlst
%Control: editor formatted (1) identically to author
%Control: production of article title (-1) disabled
%Control: page (0) single
%Control: year (1) truncated
%Control: production of eprint (0) enabled
\begin{thebibliography}{51}%
\makeatletter
\providecommand \@ifxundefined [1]{%
 \@ifx{#1\undefined}
}%
\providecommand \@ifnum [1]{%
 \ifnum #1\expandafter \@firstoftwo
 \else \expandafter \@secondoftwo
 \fi
}%
\providecommand \@ifx [1]{%
 \ifx #1\expandafter \@firstoftwo
 \else \expandafter \@secondoftwo
 \fi
}%
\providecommand \natexlab [1]{#1}%
\providecommand \enquote  [1]{``#1''}%
\providecommand \bibnamefont  [1]{#1}%
\providecommand \bibfnamefont [1]{#1}%
\providecommand \citenamefont [1]{#1}%
\providecommand \href@noop [0]{\@secondoftwo}%
\providecommand \href [0]{\begingroup \@sanitize@url \@href}%
\providecommand \@href[1]{\@@startlink{#1}\@@href}%
\providecommand \@@href[1]{\endgroup#1\@@endlink}%
\providecommand \@sanitize@url [0]{\catcode `\\12\catcode `\$12\catcode
  `\&12\catcode `\#12\catcode `\^12\catcode `\_12\catcode `\%12\relax}%
\providecommand \@@startlink[1]{}%
\providecommand \@@endlink[0]{}%
\providecommand \url  [0]{\begingroup\@sanitize@url \@url }%
\providecommand \@url [1]{\endgroup\@href {#1}{\urlprefix }}%
\providecommand \urlprefix  [0]{URL }%
\providecommand \Eprint [0]{\href }%
\providecommand \doibase [0]{http://dx.doi.org/}%
\providecommand \selectlanguage [0]{\@gobble}%
\providecommand \bibinfo  [0]{\@secondoftwo}%
\providecommand \bibfield  [0]{\@secondoftwo}%
\providecommand \translation [1]{[#1]}%
\providecommand \BibitemOpen [0]{}%
\providecommand \bibitemStop [0]{}%
\providecommand \bibitemNoStop [0]{.\EOS\space}%
\providecommand \EOS [0]{\spacefactor3000\relax}%
\providecommand \BibitemShut  [1]{\csname bibitem#1\endcsname}%
\let\auto@bib@innerbib\@empty
%</preamble>
\bibitem [{\citenamefont {Leimkuhler}\ and\ \citenamefont
  {Matthews}(2015)}]{29acd3d494044594aea0829ef236aad6}%
  \BibitemOpen
  \bibfield  {author} {\bibinfo {author} {\bibfnamefont {B.}~\bibnamefont
  {Leimkuhler}}\ and\ \bibinfo {author} {\bibfnamefont {C.}~\bibnamefont
  {Matthews}},\ }\href {\doibase 10.1007/978-3-319-16375-8} {\emph {\bibinfo
  {title} {Molecular Dynamics: With Deterministic and Stochastic Numerical
  Methods}}},\ Interdisciplinary Applied Mathematics\ (\bibinfo  {publisher}
  {Springer},\ \bibinfo {address} {Cham},\ \bibinfo {year} {2015})\BibitemShut
  {NoStop}%
\bibitem [{\citenamefont {Bussi}\ and\ \citenamefont
  {Parrinello}(2007)}]{bussi2007accurate}%
  \BibitemOpen
  \bibfield  {author} {\bibinfo {author} {\bibfnamefont {G.}~\bibnamefont
  {Bussi}}\ and\ \bibinfo {author} {\bibfnamefont {M.}~\bibnamefont
  {Parrinello}},\ }\href {\doibase 10.1103/PhysRevE.75.056707} {\bibfield
  {journal} {\bibinfo  {journal} {Physical Review E}\ }\textbf {\bibinfo
  {volume} {75}},\ \bibinfo {pages} {056707} (\bibinfo {year}
  {2007})}\BibitemShut {NoStop}%
\bibitem [{\citenamefont {Frenkel}\ and\ \citenamefont
  {Smit}(2001)}]{frenkel2001understanding}%
  \BibitemOpen
  \bibfield  {author} {\bibinfo {author} {\bibfnamefont {D.}~\bibnamefont
  {Frenkel}}\ and\ \bibinfo {author} {\bibfnamefont {B.}~\bibnamefont {Smit}},\
  }\href@noop {} {\emph {\bibinfo {title} {Understanding Molecular
  Simulation}}},\ \bibinfo {edition} {2nd}\ ed.\ (\bibinfo  {publisher}
  {Academic Press, Inc.},\ \bibinfo {address} {USA},\ \bibinfo {year}
  {2001})\BibitemShut {NoStop}%
\bibitem [{\citenamefont {Allen}\ and\ \citenamefont
  {Tildesley}(2017)}]{allen_computer_2017}%
  \BibitemOpen
  \bibfield  {author} {\bibinfo {author} {\bibfnamefont {M.~P.}\ \bibnamefont
  {Allen}}\ and\ \bibinfo {author} {\bibfnamefont {D.~J.}\ \bibnamefont
  {Tildesley}},\ }\href {\doibase 10.1093/oso/9780198803195.001.0001} {\emph
  {\bibinfo {title} {Computer {Simulation} of {Liquids}}}}\ (\bibinfo
  {publisher} {Oxford University Press},\ \bibinfo {year} {2017})\BibitemShut
  {NoStop}%
\bibitem [{\citenamefont {Pastor}(1994)}]{Pastor1994}%
  \BibitemOpen
  \bibfield  {author} {\bibinfo {author} {\bibfnamefont {R.~W.}\ \bibnamefont
  {Pastor}},\ }\enquote {\bibinfo {title} {Techniques and applications of
  {Langevin} dynamics simulations},}\ in\ \href {\doibase
  10.1007/978-94-011-1168-3_5} {\emph {\bibinfo {booktitle} {The Molecular
  Dynamics of Liquid Crystals}}},\ \bibinfo {editor} {edited by\ \bibinfo
  {editor} {\bibfnamefont {G.~R.}\ \bibnamefont {Luckhurst}}\ and\ \bibinfo
  {editor} {\bibfnamefont {C.~A.}\ \bibnamefont {Veracini}}}\ (\bibinfo
  {publisher} {Springer Netherlands},\ \bibinfo {address} {Dordrecht},\
  \bibinfo {year} {1994})\ pp.\ \bibinfo {pages} {85--138}\BibitemShut
  {NoStop}%
\bibitem [{\citenamefont {Darden}, \citenamefont {York},\ and\ \citenamefont
  {Pedersen}(1993)}]{darden_particle_1993}%
  \BibitemOpen
  \bibfield  {author} {\bibinfo {author} {\bibfnamefont {T.}~\bibnamefont
  {Darden}}, \bibinfo {author} {\bibfnamefont {D.}~\bibnamefont {York}}, \ and\
  \bibinfo {author} {\bibfnamefont {L.}~\bibnamefont {Pedersen}},\ }\href
  {\doibase 10.1063/1.464397} {\bibfield  {journal} {\bibinfo  {journal} {The
  Journal of Chemical Physics}\ }\textbf {\bibinfo {volume} {98}},\ \bibinfo
  {pages} {10089} (\bibinfo {year} {1993})}\BibitemShut {NoStop}%
\bibitem [{\citenamefont {Plimpton}(1995)}]{plimpton1995lammps}%
  \BibitemOpen
  \bibfield  {author} {\bibinfo {author} {\bibfnamefont {S.}~\bibnamefont
  {Plimpton}},\ }\href {\doibase 10.1006/jcph.1995.1039} {\bibfield  {journal}
  {\bibinfo  {journal} {Journal of Computational Physics}\ }\textbf {\bibinfo
  {volume} {117}},\ \bibinfo {pages} {1} (\bibinfo {year} {1995})}\BibitemShut
  {NoStop}%
\bibitem [{\citenamefont {Thompson}\ \emph {et~al.}(2022)\citenamefont
  {Thompson}, \citenamefont {Aktulga}, \citenamefont {Berger}, \citenamefont
  {Bolintineanu}, \citenamefont {Brown}, \citenamefont {Crozier}, \citenamefont
  {in~'t Veld}, \citenamefont {Kohlmeyer}, \citenamefont {Moore}, \citenamefont
  {Nguyen}, \citenamefont {Shan}, \citenamefont {Stevens}, \citenamefont
  {Tranchida}, \citenamefont {Trott},\ and\ \citenamefont
  {Plimpton}}]{thompson2022lammps}%
  \BibitemOpen
  \bibfield  {author} {\bibinfo {author} {\bibfnamefont {A.~P.}\ \bibnamefont
  {Thompson}}, \bibinfo {author} {\bibfnamefont {H.~M.}\ \bibnamefont
  {Aktulga}}, \bibinfo {author} {\bibfnamefont {R.}~\bibnamefont {Berger}},
  \bibinfo {author} {\bibfnamefont {D.~S.}\ \bibnamefont {Bolintineanu}},
  \bibinfo {author} {\bibfnamefont {W.~M.}\ \bibnamefont {Brown}}, \bibinfo
  {author} {\bibfnamefont {P.~S.}\ \bibnamefont {Crozier}}, \bibinfo {author}
  {\bibfnamefont {P.~J.}\ \bibnamefont {in~'t Veld}}, \bibinfo {author}
  {\bibfnamefont {A.}~\bibnamefont {Kohlmeyer}}, \bibinfo {author}
  {\bibfnamefont {S.~G.}\ \bibnamefont {Moore}}, \bibinfo {author}
  {\bibfnamefont {T.~D.}\ \bibnamefont {Nguyen}}, \bibinfo {author}
  {\bibfnamefont {R.}~\bibnamefont {Shan}}, \bibinfo {author} {\bibfnamefont
  {M.~J.}\ \bibnamefont {Stevens}}, \bibinfo {author} {\bibfnamefont
  {J.}~\bibnamefont {Tranchida}}, \bibinfo {author} {\bibfnamefont
  {C.}~\bibnamefont {Trott}}, \ and\ \bibinfo {author} {\bibfnamefont {S.~J.}\
  \bibnamefont {Plimpton}},\ }\href {\doibase 10.1016/j.cpc.2021.108171}
  {\bibfield  {journal} {\bibinfo  {journal} {Computer Physics Communications}\
  }\textbf {\bibinfo {volume} {271}},\ \bibinfo {pages} {108171} (\bibinfo
  {year} {2022})}\BibitemShut {NoStop}%
\bibitem [{\citenamefont {Sagui}\ and\ \citenamefont
  {Darden}(1999)}]{Long-RangeelectrostaticEffects}%
  \BibitemOpen
  \bibfield  {author} {\bibinfo {author} {\bibfnamefont {C.}~\bibnamefont
  {Sagui}}\ and\ \bibinfo {author} {\bibfnamefont {T.~A.}\ \bibnamefont
  {Darden}},\ }\href {\doibase 10.1146/annurev.biophys.28.1.155} {\bibfield
  {journal} {\bibinfo  {journal} {Annual Review of Biophysics and Biomolecular
  Structure}\ }\textbf {\bibinfo {volume} {28}},\ \bibinfo {pages} {155}
  (\bibinfo {year} {1999})}\BibitemShut {NoStop}%
\bibitem [{\citenamefont {Simmonett}\ and\ \citenamefont
  {Brooks}(2021)}]{simmonett_compression_2021}%
  \BibitemOpen
  \bibfield  {author} {\bibinfo {author} {\bibfnamefont {A.~C.}\ \bibnamefont
  {Simmonett}}\ and\ \bibinfo {author} {\bibfnamefont {B.~R.}\ \bibnamefont
  {Brooks}},\ }\href {\doibase 10.1063/5.0040966} {\bibfield  {journal}
  {\bibinfo  {journal} {The Journal of Chemical Physics}\ }\textbf {\bibinfo
  {volume} {154}},\ \bibinfo {pages} {054112} (\bibinfo {year}
  {2021})}\BibitemShut {NoStop}%
\bibitem [{\citenamefont {Robbins}\ and\ \citenamefont
  {Monro}(1951)}]{robbins1951misc-stochastic}%
  \BibitemOpen
  \bibfield  {author} {\bibinfo {author} {\bibfnamefont {H.}~\bibnamefont
  {Robbins}}\ and\ \bibinfo {author} {\bibfnamefont {S.}~\bibnamefont
  {Monro}},\ }\href {\doibase 10.1214/aoms/1177729586} {\bibfield  {journal}
  {\bibinfo  {journal} {Annals of Mathematical Statistics}\ }\textbf {\bibinfo
  {volume} {22}},\ \bibinfo {pages} {400} (\bibinfo {year} {1951})}\BibitemShut
  {NoStop}%
\bibitem [{\citenamefont {Bottou}(2010)}]{summa_large-scale_2010}%
  \BibitemOpen
  \bibfield  {author} {\bibinfo {author} {\bibfnamefont {L.}~\bibnamefont
  {Bottou}},\ }in\ \href {\doibase 10.1007/978-3-7908-2604-3_16} {\emph
  {\bibinfo {booktitle} {Proceedings of COMPSTAT'2010}}},\ \bibinfo {editor}
  {edited by\ \bibinfo {editor} {\bibfnamefont {Y.}~\bibnamefont
  {Lechevallier}}\ and\ \bibinfo {editor} {\bibfnamefont {G.}~\bibnamefont
  {Saporta}}}\ (\bibinfo  {publisher} {Physica-Verlag HD},\ \bibinfo {address}
  {Heidelberg},\ \bibinfo {year} {2010})\ pp.\ \bibinfo {pages}
  {177--186}\BibitemShut {NoStop}%
\bibitem [{\citenamefont {Jin}, \citenamefont {Li},\ and\ \citenamefont
  {Liu}(2020)}]{jin2020rbm}%
  \BibitemOpen
  \bibfield  {author} {\bibinfo {author} {\bibfnamefont {S.}~\bibnamefont
  {Jin}}, \bibinfo {author} {\bibfnamefont {L.}~\bibnamefont {Li}}, \ and\
  \bibinfo {author} {\bibfnamefont {J.-G.}\ \bibnamefont {Liu}},\ }\href
  {\doibase 10.1016/j.jcp.2019.108877} {\bibfield  {journal} {\bibinfo
  {journal} {Journal of Computational Physics}\ }\textbf {\bibinfo {volume}
  {400}},\ \bibinfo {pages} {108877} (\bibinfo {year} {2020})}\BibitemShut
  {NoStop}%
\bibitem [{\citenamefont {Golse}, \citenamefont {Jin},\ and\ \citenamefont
  {Paul}(2021)}]{golse:hal-02405783}%
  \BibitemOpen
  \bibfield  {author} {\bibinfo {author} {\bibfnamefont {F.}~\bibnamefont
  {Golse}}, \bibinfo {author} {\bibfnamefont {S.}~\bibnamefont {Jin}}, \ and\
  \bibinfo {author} {\bibfnamefont {T.}~\bibnamefont {Paul}},\ }\href {\doibase
  10.4208/jcm.2107-m2020-0306} {\bibfield  {journal} {\bibinfo  {journal}
  {Journal of Computational Mathematics}\ }\textbf {\bibinfo {volume} {39}},\
  \bibinfo {pages} {897} (\bibinfo {year} {2021})}\BibitemShut {NoStop}%
\bibitem [{\citenamefont {Li}, \citenamefont {Xu},\ and\ \citenamefont
  {Zhao}(2020)}]{RBMCMB}%
  \BibitemOpen
  \bibfield  {author} {\bibinfo {author} {\bibfnamefont {L.}~\bibnamefont
  {Li}}, \bibinfo {author} {\bibfnamefont {Z.}~\bibnamefont {Xu}}, \ and\
  \bibinfo {author} {\bibfnamefont {Y.}~\bibnamefont {Zhao}},\ }\href {\doibase
  10.1137/19M1302077} {\bibfield  {journal} {\bibinfo  {journal} {SIAM Journal
  on Scientific Computing}\ }\textbf {\bibinfo {volume} {42}},\ \bibinfo
  {pages} {A1486} (\bibinfo {year} {2020})}\BibitemShut {NoStop}%
\bibitem [{\citenamefont {Jin}\ and\ \citenamefont {Li}(2022)}]{jin_mean_2022}%
  \BibitemOpen
  \bibfield  {author} {\bibinfo {author} {\bibfnamefont {S.}~\bibnamefont
  {Jin}}\ and\ \bibinfo {author} {\bibfnamefont {L.}~\bibnamefont {Li}},\
  }\href {\doibase 10.1007/s11425-020-1810-6} {\bibfield  {journal} {\bibinfo
  {journal} {Science China Mathematics}\ }\textbf {\bibinfo {volume} {65}},\
  \bibinfo {pages} {169} (\bibinfo {year} {2022})}\BibitemShut {NoStop}%
\bibitem [{\citenamefont {Ko}\ and\ \citenamefont
  {Zuazua}(2021)}]{RB_on_guiding_problem}%
  \BibitemOpen
  \bibfield  {author} {\bibinfo {author} {\bibfnamefont {D.}~\bibnamefont
  {Ko}}\ and\ \bibinfo {author} {\bibfnamefont {E.}~\bibnamefont {Zuazua}},\
  }\href {\doibase 10.1142/S0218202521500329} {\bibfield  {journal} {\bibinfo
  {journal} {Mathematical Models and Methods in Applied Sciences}\ }\textbf
  {\bibinfo {volume} {31}},\ \bibinfo {pages} {1569} (\bibinfo {year}
  {2021})}\BibitemShut {NoStop}%
\bibitem [{\citenamefont {Ye}\ and\ \citenamefont
  {Zhou}(2024)}]{ye_error_2024}%
  \BibitemOpen
  \bibfield  {author} {\bibinfo {author} {\bibfnamefont {X.}~\bibnamefont
  {Ye}}\ and\ \bibinfo {author} {\bibfnamefont {Z.}~\bibnamefont {Zhou}},\
  }\href {\doibase 10.1093/imanum/drad043} {\bibfield  {journal} {\bibinfo
  {journal} {IMA Journal of Numerical Analysis}\ }\textbf {\bibinfo {volume}
  {44}},\ \bibinfo {pages} {1660} (\bibinfo {year} {2024})}\BibitemShut
  {NoStop}%
\bibitem [{\citenamefont {Cai}, \citenamefont {Liu},\ and\ \citenamefont
  {Wang}(2026)}]{Cai2024ConvergenceOR}%
  \BibitemOpen
  \bibfield  {author} {\bibinfo {author} {\bibfnamefont {Z.}~\bibnamefont
  {Cai}}, \bibinfo {author} {\bibfnamefont {J.-G.}\ \bibnamefont {Liu}}, \ and\
  \bibinfo {author} {\bibfnamefont {Y.}~\bibnamefont {Wang}},\ }\href {\doibase
  10.1090/mcom/4187} {\bibfield  {journal} {\bibinfo  {journal} {Mathematics of
  Computation}\ } (\bibinfo {year} {2026}),\ 10.1090/mcom/4187}\BibitemShut
  {NoStop}%
\bibitem [{\citenamefont {Jin}\ \emph {et~al.}(2021)\citenamefont {Jin},
  \citenamefont {Li}, \citenamefont {Xu},\ and\ \citenamefont
  {Zhao}}]{jin2021rbe}%
  \BibitemOpen
  \bibfield  {author} {\bibinfo {author} {\bibfnamefont {S.}~\bibnamefont
  {Jin}}, \bibinfo {author} {\bibfnamefont {L.}~\bibnamefont {Li}}, \bibinfo
  {author} {\bibfnamefont {Z.}~\bibnamefont {Xu}}, \ and\ \bibinfo {author}
  {\bibfnamefont {Y.}~\bibnamefont {Zhao}},\ }\href {\doibase
  10.1137/20M1371385} {\bibfield  {journal} {\bibinfo  {journal} {SIAM Journal
  on Scientific Computing}\ }\textbf {\bibinfo {volume} {43}},\ \bibinfo
  {pages} {B937} (\bibinfo {year} {2021})}\BibitemShut {NoStop}%
\bibitem [{\citenamefont {Liang}\ \emph
  {et~al.}(2022{\natexlab{a}})\citenamefont {Liang}, \citenamefont {Tan},
  \citenamefont {Zhao}, \citenamefont {Li}, \citenamefont {Jin}, \citenamefont
  {Hong},\ and\ \citenamefont {Xu}}]{liang_superscalability_2022}%
  \BibitemOpen
  \bibfield  {author} {\bibinfo {author} {\bibfnamefont {J.}~\bibnamefont
  {Liang}}, \bibinfo {author} {\bibfnamefont {P.}~\bibnamefont {Tan}}, \bibinfo
  {author} {\bibfnamefont {Y.}~\bibnamefont {Zhao}}, \bibinfo {author}
  {\bibfnamefont {L.}~\bibnamefont {Li}}, \bibinfo {author} {\bibfnamefont
  {S.}~\bibnamefont {Jin}}, \bibinfo {author} {\bibfnamefont {L.}~\bibnamefont
  {Hong}}, \ and\ \bibinfo {author} {\bibfnamefont {Z.}~\bibnamefont {Xu}},\
  }\href {\doibase 10.1063/5.0073424} {\bibfield  {journal} {\bibinfo
  {journal} {The Journal of Chemical Physics}\ }\textbf {\bibinfo {volume}
  {156}},\ \bibinfo {pages} {014114} (\bibinfo {year}
  {2022}{\natexlab{a}})}\BibitemShut {NoStop}%
\bibitem [{\citenamefont {Liang}\ \emph
  {et~al.}(2022{\natexlab{b}})\citenamefont {Liang}, \citenamefont {Tan},
  \citenamefont {Hong}, \citenamefont {Jin}, \citenamefont {Xu},\ and\
  \citenamefont {Li}}]{liang_random_2022_npt}%
  \BibitemOpen
  \bibfield  {author} {\bibinfo {author} {\bibfnamefont {J.}~\bibnamefont
  {Liang}}, \bibinfo {author} {\bibfnamefont {P.}~\bibnamefont {Tan}}, \bibinfo
  {author} {\bibfnamefont {L.}~\bibnamefont {Hong}}, \bibinfo {author}
  {\bibfnamefont {S.}~\bibnamefont {Jin}}, \bibinfo {author} {\bibfnamefont
  {Z.}~\bibnamefont {Xu}}, \ and\ \bibinfo {author} {\bibfnamefont
  {L.}~\bibnamefont {Li}},\ }\href {\doibase 10.1063/5.0107140} {\bibfield
  {journal} {\bibinfo  {journal} {The Journal of Chemical Physics}\ }\textbf
  {\bibinfo {volume} {157}},\ \bibinfo {pages} {144102} (\bibinfo {year}
  {2022}{\natexlab{b}})}\BibitemShut {NoStop}%
\bibitem [{\citenamefont {Liang}, \citenamefont {Xu},\ and\ \citenamefont
  {Zhou}(2023)}]{liang2023rbsog}%
  \BibitemOpen
  \bibfield  {author} {\bibinfo {author} {\bibfnamefont {J.}~\bibnamefont
  {Liang}}, \bibinfo {author} {\bibfnamefont {Z.}~\bibnamefont {Xu}}, \ and\
  \bibinfo {author} {\bibfnamefont {Q.}~\bibnamefont {Zhou}},\ }\href {\doibase
  10.1137/22M1497201} {\bibfield  {journal} {\bibinfo  {journal} {SIAM Journal
  on Scientific Computing}\ }\textbf {\bibinfo {volume} {45}},\ \bibinfo
  {pages} {B591} (\bibinfo {year} {2023})}\BibitemShut {NoStop}%
\bibitem [{\citenamefont {Gan}\ \emph {et~al.}(2025)\citenamefont {Gan},
  \citenamefont {Gao}, \citenamefont {Liang},\ and\ \citenamefont
  {Xu}}]{quasi-2DRBE}%
  \BibitemOpen
  \bibfield  {author} {\bibinfo {author} {\bibfnamefont {Z.}~\bibnamefont
  {Gan}}, \bibinfo {author} {\bibfnamefont {X.}~\bibnamefont {Gao}}, \bibinfo
  {author} {\bibfnamefont {J.}~\bibnamefont {Liang}}, \ and\ \bibinfo {author}
  {\bibfnamefont {Z.}~\bibnamefont {Xu}},\ }\href {\doibase 10.1137/24M1655809}
  {\bibfield  {journal} {\bibinfo  {journal} {SIAM Journal on Scientific
  Computing}\ }\textbf {\bibinfo {volume} {47}},\ \bibinfo {pages} {B846}
  (\bibinfo {year} {2025})}\BibitemShut {NoStop}%
\bibitem [{\citenamefont {Liang}, \citenamefont {Xu},\ and\ \citenamefont
  {Zhao}(2021)}]{liang2021rbl}%
  \BibitemOpen
  \bibfield  {author} {\bibinfo {author} {\bibfnamefont {J.}~\bibnamefont
  {Liang}}, \bibinfo {author} {\bibfnamefont {Z.}~\bibnamefont {Xu}}, \ and\
  \bibinfo {author} {\bibfnamefont {Y.}~\bibnamefont {Zhao}},\ }\href {\doibase
  10.1063/5.0056515} {\bibfield  {journal} {\bibinfo  {journal} {The Journal of
  Chemical Physics}\ }\textbf {\bibinfo {volume} {155}},\ \bibinfo {pages}
  {044108} (\bibinfo {year} {2021})}\BibitemShut {NoStop}%
\bibitem [{\citenamefont {Liang}, \citenamefont {Xu},\ and\ \citenamefont
  {Zhao}(2022)}]{liang_improved_2022}%
  \BibitemOpen
  \bibfield  {author} {\bibinfo {author} {\bibfnamefont {J.}~\bibnamefont
  {Liang}}, \bibinfo {author} {\bibfnamefont {Z.}~\bibnamefont {Xu}}, \ and\
  \bibinfo {author} {\bibfnamefont {Y.}~\bibnamefont {Zhao}},\ }\href {\doibase
  10.1021/acs.jpca.2c01918} {\bibfield  {journal} {\bibinfo  {journal} {The
  Journal of Physical Chemistry A}\ }\textbf {\bibinfo {volume} {126}},\
  \bibinfo {pages} {3583} (\bibinfo {year} {2022})}\BibitemShut {NoStop}%
\bibitem [{\citenamefont {Zhang}, \citenamefont {Huang},\ and\ \citenamefont
  {Yang}(2026)}]{ZHANG2026110170}%
  \BibitemOpen
  \bibfield  {author} {\bibinfo {author} {\bibfnamefont {J.}~\bibnamefont
  {Zhang}}, \bibinfo {author} {\bibfnamefont {J.}~\bibnamefont {Huang}}, \ and\
  \bibinfo {author} {\bibfnamefont {Z.}~\bibnamefont {Yang}},\ }\href {\doibase
  10.1016/j.cpc.2026.110170} {\bibfield  {journal} {\bibinfo  {journal}
  {Computer Physics Communications}\ }\textbf {\bibinfo {volume} {325}},\
  \bibinfo {pages} {110170} (\bibinfo {year} {2026})}\BibitemShut {NoStop}%
\bibitem [{\citenamefont {Gao}\ \emph {et~al.}(2026)\citenamefont {Gao},
  \citenamefont {Zhao}, \citenamefont {Guo}, \citenamefont {Liang},
  \citenamefont {Liu}, \citenamefont {Luo}, \citenamefont {Luo}, \citenamefont
  {Qin}, \citenamefont {Wang}, \citenamefont {Zhou}, \citenamefont {Jin},\ and\
  \citenamefont {Xu}}]{gao_rbmd_2025}%
  \BibitemOpen
  \bibfield  {author} {\bibinfo {author} {\bibfnamefont {W.}~\bibnamefont
  {Gao}}, \bibinfo {author} {\bibfnamefont {T.}~\bibnamefont {Zhao}}, \bibinfo
  {author} {\bibfnamefont {Y.}~\bibnamefont {Guo}}, \bibinfo {author}
  {\bibfnamefont {J.}~\bibnamefont {Liang}}, \bibinfo {author} {\bibfnamefont
  {H.}~\bibnamefont {Liu}}, \bibinfo {author} {\bibfnamefont {M.}~\bibnamefont
  {Luo}}, \bibinfo {author} {\bibfnamefont {Z.}~\bibnamefont {Luo}}, \bibinfo
  {author} {\bibfnamefont {W.}~\bibnamefont {Qin}}, \bibinfo {author}
  {\bibfnamefont {Y.}~\bibnamefont {Wang}}, \bibinfo {author} {\bibfnamefont
  {Q.}~\bibnamefont {Zhou}}, \bibinfo {author} {\bibfnamefont {S.}~\bibnamefont
  {Jin}}, \ and\ \bibinfo {author} {\bibfnamefont {Z.}~\bibnamefont {Xu}},\
  }\href {\doibase 10.4208/cicp.OA-2024-0156} {\bibfield  {journal} {\bibinfo
  {journal} {Communications in Computational Physics}\ }\textbf {\bibinfo
  {volume} {39}},\ \bibinfo {pages} {296} (\bibinfo {year} {2026})}\BibitemShut
  {NoStop}%
\bibitem [{\citenamefont {Liang}, \citenamefont {Xu},\ and\ \citenamefont
  {Zhao}(2024)}]{liang2024ess}%
  \BibitemOpen
  \bibfield  {author} {\bibinfo {author} {\bibfnamefont {J.}~\bibnamefont
  {Liang}}, \bibinfo {author} {\bibfnamefont {Z.}~\bibnamefont {Xu}}, \ and\
  \bibinfo {author} {\bibfnamefont {Y.}~\bibnamefont {Zhao}},\ }\href {\doibase
  10.1063/5.0187108} {\bibfield  {journal} {\bibinfo  {journal} {The Journal of
  Chemical Physics}\ }\textbf {\bibinfo {volume} {160}},\ \bibinfo {pages}
  {034101} (\bibinfo {year} {2024})}\BibitemShut {NoStop}%
\bibitem [{\citenamefont {Kubo}(1966)}]{Kubo}%
  \BibitemOpen
  \bibfield  {author} {\bibinfo {author} {\bibfnamefont {R.}~\bibnamefont
  {Kubo}},\ }\href {\doibase 10.1088/0034-4885/29/1/306} {\bibfield  {journal}
  {\bibinfo  {journal} {Reports on Progress in Physics}\ }\textbf {\bibinfo
  {volume} {29}},\ \bibinfo {pages} {255} (\bibinfo {year} {1966})}\BibitemShut
  {NoStop}%
\bibitem [{\citenamefont {Xu}, \citenamefont {Zhao},\ and\ \citenamefont
  {Zhou}(2024)}]{xu2024vrrbl}%
  \BibitemOpen
  \bibfield  {author} {\bibinfo {author} {\bibfnamefont {Z.}~\bibnamefont
  {Xu}}, \bibinfo {author} {\bibfnamefont {Y.}~\bibnamefont {Zhao}}, \ and\
  \bibinfo {author} {\bibfnamefont {Q.}~\bibnamefont {Zhou}},\ }\href {\doibase
  10.1063/5.0246661} {\bibfield  {journal} {\bibinfo  {journal} {The Journal of
  Chemical Physics}\ }\textbf {\bibinfo {volume} {161}},\ \bibinfo {pages}
  {244110} (\bibinfo {year} {2024})}\BibitemShut {NoStop}%
\bibitem [{\citenamefont {Jin}, \citenamefont {Li},\ and\ \citenamefont
  {Sun}(2022)}]{jin_random_2020}%
  \BibitemOpen
  \bibfield  {author} {\bibinfo {author} {\bibfnamefont {S.}~\bibnamefont
  {Jin}}, \bibinfo {author} {\bibfnamefont {L.}~\bibnamefont {Li}}, \ and\
  \bibinfo {author} {\bibfnamefont {Y.}~\bibnamefont {Sun}},\ }\href {\doibase
  10.1137/20M1383069} {\bibfield  {journal} {\bibinfo  {journal} {Multiscale
  Modeling \& Simulation}\ }\textbf {\bibinfo {volume} {20}},\ \bibinfo {pages}
  {741} (\bibinfo {year} {2022})}\BibitemShut {NoStop}%
\bibitem [{\citenamefont {Basconi}\ and\ \citenamefont
  {Shirts}(2013)}]{basconi_effects_2013}%
  \BibitemOpen
  \bibfield  {author} {\bibinfo {author} {\bibfnamefont {J.~E.}\ \bibnamefont
  {Basconi}}\ and\ \bibinfo {author} {\bibfnamefont {M.~R.}\ \bibnamefont
  {Shirts}},\ }\href {\doibase 10.1021/ct400109a} {\bibfield  {journal}
  {\bibinfo  {journal} {Journal of Chemical Theory and Computation}\ }\textbf
  {\bibinfo {volume} {9}},\ \bibinfo {pages} {2887} (\bibinfo {year}
  {2013})}\BibitemShut {NoStop}%
\bibitem [{\citenamefont {Leimkuhler}, \citenamefont {Matthews},\ and\
  \citenamefont {Stoltz}(2016)}]{leimkuhlercomputation2016}%
  \BibitemOpen
  \bibfield  {author} {\bibinfo {author} {\bibfnamefont {B.}~\bibnamefont
  {Leimkuhler}}, \bibinfo {author} {\bibfnamefont {C.}~\bibnamefont
  {Matthews}}, \ and\ \bibinfo {author} {\bibfnamefont {G.}~\bibnamefont
  {Stoltz}},\ }\href {\doibase 10.1093/imanum/dru056} {\bibfield  {journal}
  {\bibinfo  {journal} {IMA Journal of Numerical Analysis}\ }\textbf {\bibinfo
  {volume} {36}},\ \bibinfo {pages} {13} (\bibinfo {year} {2016})}\BibitemShut
  {NoStop}%
\bibitem [{\citenamefont {Welling}\ and\ \citenamefont
  {Teh}(2011)}]{welling2011icml-bayesian}%
  \BibitemOpen
  \bibfield  {author} {\bibinfo {author} {\bibfnamefont {M.}~\bibnamefont
  {Welling}}\ and\ \bibinfo {author} {\bibfnamefont {Y.~W.}\ \bibnamefont
  {Teh}},\ }in\ \href
  {https://mlanthology.org/icml/2011/welling2011icml-bayesian/} {\emph
  {\bibinfo {booktitle} {International Conference on Machine Learning}}}\
  (\bibinfo {year} {2011})\ pp.\ \bibinfo {pages} {681--688}\BibitemShut
  {NoStop}%
\bibitem [{\citenamefont {Kingma}\ and\ \citenamefont
  {Ba}(2015)}]{kingma2015adam}%
  \BibitemOpen
  \bibfield  {author} {\bibinfo {author} {\bibfnamefont {D.~P.}\ \bibnamefont
  {Kingma}}\ and\ \bibinfo {author} {\bibfnamefont {J.}~\bibnamefont {Ba}},\
  }in\ \href {https://arxiv.org/abs/1412.6980} {\emph {\bibinfo {booktitle}
  {International Conference on Learning Representations}}}\ (\bibinfo {year}
  {2015})\BibitemShut {NoStop}%
\bibitem [{\citenamefont {Reddi}, \citenamefont {Kale},\ and\ \citenamefont
  {Kumar}(2018)}]{reddi2018adam}%
  \BibitemOpen
  \bibfield  {author} {\bibinfo {author} {\bibfnamefont {S.}~\bibnamefont
  {Reddi}}, \bibinfo {author} {\bibfnamefont {S.}~\bibnamefont {Kale}}, \ and\
  \bibinfo {author} {\bibfnamefont {S.}~\bibnamefont {Kumar}},\ }in\ \href@noop
  {} {\emph {\bibinfo {booktitle} {International Conference on Learning
  Representations}}}\ (\bibinfo {year} {2018})\BibitemShut {NoStop}%
\bibitem [{\citenamefont {Ewald}(1921)}]{ewald1921berechnung}%
  \BibitemOpen
  \bibfield  {author} {\bibinfo {author} {\bibfnamefont {P.~P.}\ \bibnamefont
  {Ewald}},\ }\href {\doibase 10.1002/andp.19213690304} {\bibfield  {journal}
  {\bibinfo  {journal} {Annalen der Physik}\ }\textbf {\bibinfo {volume}
  {369}},\ \bibinfo {pages} {253} (\bibinfo {year} {1921})}\BibitemShut
  {NoStop}%
\bibitem [{\citenamefont {Hu}(2014)}]{hu2014infinite}%
  \BibitemOpen
  \bibfield  {author} {\bibinfo {author} {\bibfnamefont {Z.}~\bibnamefont
  {Hu}},\ }\href {\doibase 10.1021/ct500704m} {\bibfield  {journal} {\bibinfo
  {journal} {Journal of Chemical Theory and Computation}\ }\textbf {\bibinfo
  {volume} {10}},\ \bibinfo {pages} {5254} (\bibinfo {year}
  {2014})}\BibitemShut {NoStop}%
\bibitem [{\citenamefont {Hockney}\ and\ \citenamefont
  {Eastwood}(1988)}]{hockney1988computer}%
  \BibitemOpen
  \bibfield  {author} {\bibinfo {author} {\bibfnamefont {R.~W.}\ \bibnamefont
  {Hockney}}\ and\ \bibinfo {author} {\bibfnamefont {J.~W.}\ \bibnamefont
  {Eastwood}},\ }\href@noop {} {\emph {\bibinfo {title} {Computer Simulation
  Using Particles}}}\ (\bibinfo  {publisher} {CRC Press},\ \bibinfo {year}
  {1988})\BibitemShut {NoStop}%
\bibitem [{\citenamefont {Essmann}\ \emph {et~al.}(1995)\citenamefont
  {Essmann}, \citenamefont {Perera}, \citenamefont {Berkowitz}, \citenamefont
  {Darden}, \citenamefont {Lee},\ and\ \citenamefont
  {Pedersen}}]{essmann_smooth_1995}%
  \BibitemOpen
  \bibfield  {author} {\bibinfo {author} {\bibfnamefont {U.}~\bibnamefont
  {Essmann}}, \bibinfo {author} {\bibfnamefont {L.}~\bibnamefont {Perera}},
  \bibinfo {author} {\bibfnamefont {M.~L.}\ \bibnamefont {Berkowitz}}, \bibinfo
  {author} {\bibfnamefont {T.}~\bibnamefont {Darden}}, \bibinfo {author}
  {\bibfnamefont {H.}~\bibnamefont {Lee}}, \ and\ \bibinfo {author}
  {\bibfnamefont {L.~G.}\ \bibnamefont {Pedersen}},\ }\href {\doibase
  10.1063/1.470117} {\bibfield  {journal} {\bibinfo  {journal} {The Journal of
  Chemical Physics}\ }\textbf {\bibinfo {volume} {103}},\ \bibinfo {pages}
  {8577} (\bibinfo {year} {1995})}\BibitemShut {NoStop}%
\bibitem [{\citenamefont {Villani}(2009)}]{villani2009optimal}%
  \BibitemOpen
  \bibfield  {author} {\bibinfo {author} {\bibfnamefont {C.}~\bibnamefont
  {Villani}},\ }\href {\doibase 10.1007/978-3-540-71050-9} {\emph {\bibinfo
  {title} {Optimal Transport: Old and New}}},\ \bibinfo {series} {Grundlehren
  der mathematischen Wissenschaften}, Vol.\ \bibinfo {volume} {338}\ (\bibinfo
  {publisher} {Springer},\ \bibinfo {address} {Berlin, Heidelberg},\ \bibinfo
  {year} {2009})\BibitemShut {NoStop}%
\bibitem [{\citenamefont {Verlet}(1967)}]{verlet1967computer}%
  \BibitemOpen
  \bibfield  {author} {\bibinfo {author} {\bibfnamefont {L.}~\bibnamefont
  {Verlet}},\ }\href {\doibase 10.1103/PhysRev.159.98} {\bibfield  {journal}
  {\bibinfo  {journal} {Phys. Rev.}\ }\textbf {\bibinfo {volume} {159}},\
  \bibinfo {pages} {98} (\bibinfo {year} {1967})}\BibitemShut {NoStop}%
\bibitem [{\citenamefont {Watanabe}, \citenamefont {Ito},\ and\ \citenamefont
  {Hu}(2012)}]{watanabe2012phase}%
  \BibitemOpen
  \bibfield  {author} {\bibinfo {author} {\bibfnamefont {H.}~\bibnamefont
  {Watanabe}}, \bibinfo {author} {\bibfnamefont {N.}~\bibnamefont {Ito}}, \
  and\ \bibinfo {author} {\bibfnamefont {C.-K.}\ \bibnamefont {Hu}},\ }\href
  {\doibase 10.1063/1.4720089} {\bibfield  {journal} {\bibinfo  {journal} {The
  Journal of Chemical Physics}\ }\textbf {\bibinfo {volume} {136}},\ \bibinfo
  {pages} {204102} (\bibinfo {year} {2012})}\BibitemShut {NoStop}%
\bibitem [{\citenamefont {Kob}\ and\ \citenamefont
  {Andersen}(1995)}]{kob1995testing}%
  \BibitemOpen
  \bibfield  {author} {\bibinfo {author} {\bibfnamefont {W.}~\bibnamefont
  {Kob}}\ and\ \bibinfo {author} {\bibfnamefont {H.~C.}\ \bibnamefont
  {Andersen}},\ }\href {\doibase 10.1103/PhysRevE.51.4626} {\bibfield
  {journal} {\bibinfo  {journal} {Phys. Rev. E}\ }\textbf {\bibinfo {volume}
  {51}},\ \bibinfo {pages} {4626} (\bibinfo {year} {1995})}\BibitemShut
  {NoStop}%
\bibitem [{\citenamefont {Malins}\ \emph {et~al.}(2013)\citenamefont {Malins},
  \citenamefont {Eggers}, \citenamefont {Tanaka},\ and\ \citenamefont
  {Royall}}]{malins_lifetimes_2013}%
  \BibitemOpen
  \bibfield  {author} {\bibinfo {author} {\bibfnamefont {A.}~\bibnamefont
  {Malins}}, \bibinfo {author} {\bibfnamefont {J.}~\bibnamefont {Eggers}},
  \bibinfo {author} {\bibfnamefont {H.}~\bibnamefont {Tanaka}}, \ and\ \bibinfo
  {author} {\bibfnamefont {C.~P.}\ \bibnamefont {Royall}},\ }\href {\doibase
  10.1039/c3fd00078h} {\bibfield  {journal} {\bibinfo  {journal} {Faraday
  Discussions}\ }\textbf {\bibinfo {volume} {167}},\ \bibinfo {pages} {405}
  (\bibinfo {year} {2013})}\BibitemShut {NoStop}%
\bibitem [{\citenamefont {Liang}, \citenamefont {Yuan},\ and\ \citenamefont
  {Xu}(2022)}]{LIANG2022108332}%
  \BibitemOpen
  \bibfield  {author} {\bibinfo {author} {\bibfnamefont {J.}~\bibnamefont
  {Liang}}, \bibinfo {author} {\bibfnamefont {J.}~\bibnamefont {Yuan}}, \ and\
  \bibinfo {author} {\bibfnamefont {Z.}~\bibnamefont {Xu}},\ }\href {\doibase
  10.1016/j.cpc.2022.108332} {\bibfield  {journal} {\bibinfo  {journal}
  {Computer Physics Communications}\ }\textbf {\bibinfo {volume} {276}},\
  \bibinfo {pages} {108332} (\bibinfo {year} {2022})}\BibitemShut {NoStop}%
\bibitem [{\citenamefont {Deserno}\ and\ \citenamefont
  {Holm}(1998)}]{deserno_how_1998}%
  \BibitemOpen
  \bibfield  {author} {\bibinfo {author} {\bibfnamefont {M.}~\bibnamefont
  {Deserno}}\ and\ \bibinfo {author} {\bibfnamefont {C.}~\bibnamefont {Holm}},\
  }\href {\doibase 10.1063/1.477415} {\bibfield  {journal} {\bibinfo  {journal}
  {The Journal of Chemical Physics}\ }\textbf {\bibinfo {volume} {109}},\
  \bibinfo {pages} {7694} (\bibinfo {year} {1998})}\BibitemShut {NoStop}%
\bibitem [{\citenamefont {Forsman}, \citenamefont {Ribar},\ and\ \citenamefont
  {Woodward}(2024)}]{forsman_efficient_2024}%
  \BibitemOpen
  \bibfield  {author} {\bibinfo {author} {\bibfnamefont {J.}~\bibnamefont
  {Forsman}}, \bibinfo {author} {\bibfnamefont {D.}~\bibnamefont {Ribar}}, \
  and\ \bibinfo {author} {\bibfnamefont {C.~E.}\ \bibnamefont {Woodward}},\
  }\href {\doibase 10.1039/D4CP00546E} {\bibfield  {journal} {\bibinfo
  {journal} {Physical Chemistry Chemical Physics}\ }\textbf {\bibinfo {volume}
  {26}},\ \bibinfo {pages} {19921} (\bibinfo {year} {2024})}\BibitemShut
  {NoStop}%
\bibitem [{\citenamefont {Kjellander}(2019)}]{kjellander2019dielectric}%
  \BibitemOpen
  \bibfield  {author} {\bibinfo {author} {\bibfnamefont {R.}~\bibnamefont
  {Kjellander}},\ }\href {\doibase 10.1039/C9SM00712A} {\bibfield  {journal}
  {\bibinfo  {journal} {Soft Matter}\ }\textbf {\bibinfo {volume} {15}},\
  \bibinfo {pages} {5866} (\bibinfo {year} {2019})}\BibitemShut {NoStop}%
\bibitem [{\citenamefont {Kjellander}(2020)}]{kjellander2020debye}%
  \BibitemOpen
  \bibfield  {author} {\bibinfo {author} {\bibfnamefont {R.}~\bibnamefont
  {Kjellander}},\ }\href {\doibase 10.1039/D0CP02742A} {\bibfield  {journal}
  {\bibinfo  {journal} {Physical Chemistry Chemical Physics}\ }\textbf
  {\bibinfo {volume} {22}},\ \bibinfo {pages} {23952} (\bibinfo {year}
  {2020})}\BibitemShut {NoStop}%
\end{thebibliography}%

\end{document}